\documentclass[%
 reprint,
 superscriptaddress, 
 amsmath,amssymb,
 aps,
 pra,
 floatfix,
]{revtex4-2}

\usepackage{graphicx}
\usepackage{dcolumn}
\usepackage{bm}
\usepackage{dsfont}
\usepackage{xcolor}
\usepackage{enumitem}
\usepackage{multirow}
\usepackage{hyperref}
\usepackage{array}
\usepackage{csquotes}
\usepackage[T1]{fontenc}
\MakeOuterQuote{"}

 \usepackage{amsmath}
    \usepackage{amssymb}
    \usepackage{amsthm}
    \usepackage{amsfonts}
    \usepackage{braket}
    \usepackage{xfrac}

\newtheorem{definition}{Definition}

\theoremstyle{remark}

\usepackage{tabularx}

\usepackage{booktabs}

\begin{document}

\title{Efficient Graph State Purification with Factorized Graph-Preserving Operations across Local Clifford Orbits}

\author{Mingyuan Wang}
\email{mingyuanwang@umass.edu}
\affiliation{Department of Physics, University of Massachusetts Amherst, Amherst, Massachusetts 01003-9337, USA}
\author{Guus Avis}
\affiliation{Manning College of Information and Computer Sciences, University of Massachusetts Amherst, Amherst, Massachusetts 01003-9264, USA}
\author{Kenneth Goodenough}
\affiliation{Naturwissenschaftlich-Technische Fakult\"{a}t, Universit\"{a}t Siegen, Walter-Flex-Stra\ss e 3, 57068 Siegen, Germany}
\author{Stefan Krastanov}
\email{skrastanov@umass.edu}
\affiliation{Department of Physics, University of Massachusetts Amherst, Amherst, Massachusetts 01003-9337, USA}
\affiliation{Manning College of Information and Computer Sciences, University of Massachusetts Amherst, Amherst, Massachusetts 01003-9264, USA}

\date{\today}
\begin{abstract}
Graph states form a broad class of multipartite entangled states underlying measurement-based quantum computation, quantum networks, and stabilizer codes. However, systematic entanglement distillation for arbitrary graph states remains challenging because the circuit design space grows rapidly with the number of parties.
We introduce a group of Clifford operations that we call "factorized graph-preserving". It enables us to efficiently enumerate and optimize graph-state purification circuits at finite size for realistic noisy hardware. These operations map products of graph-basis states to products of graph-basis states, so their action can be represented as permutations of graph-basis labels. Moreover, this useful gate set admits a compact factorized description determined by simple graph-theoretic features. This structure also allows, after some initial cached precomputation, drastically lower computational complexity for simulating a gate.
We further organize these operations over local-complementation (LC) orbits using minimum-edge representatives (MERs), which let us design purification circuits that apply to all locally equivalent graph states (up to a basis change). Using this framework, we optimize noisy finite-size multipartite distillation circuits for several graph-state families. Numerical results show that the resulting graph-preserving circuits can outperform standard recurrence-based purification protocols under realistic gate and measurement noise. Our results establish LC-orbit structure and factorized graph-preserving operations as practical tools for scalable, topology-aware and hardware-constrained graph-state distillation protocol design.
Our work can also be interpreted as a graph-based heuristic for finding transversal gates.

\end{abstract}

\maketitle

\section{Introduction}
Graph states form a broad and expressive family of multipartite entangled states whose correlations are naturally captured by an underlying graph \cite{Hein2006GraphStates,Dur2007Review}. 
Each vertex represents a qubit and each edge corresponds to an entangling interaction, providing a compact description of how entanglement is distributed across a quantum system. 
Because of this representation, graph states serve as a central resource in many areas of quantum information processing, including measurement-based quantum computation \cite{raussendorf2001oneway,raussendorf2003mbqc,briegel2009mbqc,kaldenbach2024mapping,vijayan2024compilation,kaldenbach2025resource}, loss-tolerant quantum error-correcting codes \cite{bell2023graphcodes,Schlingemann2001GraphCodes}, and distributed quantum networks \cite{Bennett1996Purification, borregaard2020repeater,Briegel1998Repeater}. 
At the same time, different graphs correspond to different multipartite correlation structures, since the edge set determines how correlations are distributed among the qubits (up to graph transformations called "local complementations"). This graph dependence makes purification of general graph states challenging. 
Although purification protocols have been developed for two-colorable graph states and extended to all graph states \cite{Dur2003TwoColorable,Kruszynska2006AllGraph}, these works do not provide a systematic framework for optimizing finite-size circuits under hardware-specific gate and measurement noise.
A circuit-level framework for identifying useful transformations and optimizing noisy distillation circuits for general graph states remains lacking.
Developing efficient methods for distilling entanglement in general graph states is therefore a key step toward scalable quantum technologies.

Our previous work has shown that for highly symmetric entangled states, such as the Greenberger-Horne-Zeilinger (GHZ) states, local Clifford transformations that map GHZ states to themselves up to Pauli corrections exhibit a straightforward group structure.
We used this notion of local operations that "map one or more GHZ states to themselves up to a Pauli correction" to search for efficient error detection (i.e.\ purification) circuits. We call such operations "GHZ-preserving". In particular, the symmetry of GHZ-preserving circuits enabled us to efficiently enumerate and simulate them, and has led to the discovery of the highest-performing noisy GHZ purification circuits~\cite{GHZPreserving}. 

For general graph states, however, the analogous Clifford operations are no longer determined by a single uniform group of symmetries. 
They depend on the stabilizer structure of the underlying graph state: different graphs can admit different local Clifford transformations that preserve the relevant graph-state structure. 

In this work, we develop a framework for classifying and analyzing a set of Clifford operations that preserve the graph-state basis of arbitrary graph states. 
Here the graph-state basis refers to the set of states obtained from a given graph state by local Pauli operations, which forms a natural basis of stabilizer states associated with the graph.
Restricting attention to operations that map graph-basis states to other graph-basis states ensures that the action of a circuit is both sufficiently general for error detection and, as we will show, much faster to simulate than general Clifford gates. Previously considered circuits~\cite{Dur2003TwoColorable,Kruszynska2006AllGraph} for graph-state distillation are a subset of this family of circuits.
This restriction dramatically reduces the operational search space: instead of arbitrary Clifford dynamics, circuit actions reduce to permutations within the discrete set of graph-basis states (only permutations that can be realized by local gates).
We show that the resulting set of admissible operations forms a finite group, whose elements and their action on states can be efficiently enumerated and simulated by tracking permutations over the graph-state basis, much faster than the typical Clifford circuit formalism~\cite{Gottesman1997Stabilizer,Aaronson2004ImprovedSimulation}. 
This group-based description enables exact and highly efficient simulation of circuit actions, making it feasible to apply large-scale numerical optimization techniques even for multipartite systems with many qubits.

The efficiency gain comes from both the graph-basis permutation picture and the way the admissible operations are represented. Rather than storing each candidate operation as an expanded Clifford circuit or a full stabilizer tableau on the two copies, we use a factorized description built from a small set of gates. For a given graph, the action of these components on graph-basis labels can be precomputed and then reused throughout the circuit search. A gate in an optimized circuit is therefore stored as a compact symbolic instruction, or equivalently as a pointer to a cached graph-basis update rule. This shifts the costly stabilizer-level analysis into a preprocessing step and makes repeated circuit evaluation substantially cheaper during optimization.

We combine this faster simulation method with circuit-optimization techniques to design entanglement-distillation circuits for general graph states under realistic noise models. 
By restricting the search to our set of graph-preserving Clifford operations and exploiting their succinct parameterization, the optimization can explore a large space of candidate circuits while remaining computationally tractable. 
This approach enables systematic optimization beyond analytically tractable protocols and allows us to study distillation performance for multipartite graph states relevant to quantum networks and distributed quantum information processing.

The remainder of this paper is organized as follows. 
Section~\ref{sec2} introduces the graph-state basis and formally defines the class of graph-preserving Clifford operations. 
We then characterize their algebraic structure and develop the simulation framework used throughout this work. 
Section~\ref{sec3} analyzes local equivalence of graph states and introduces the minimum-edge representative (MER), which provides a canonical reference for determining the admissible gate set within each LC orbit. 
Section~\ref{sec4} applies the resulting framework to the design and optimization of graph-state distillation circuits under realistic noise models. 
We conclude in Section~\ref{sec5} with a discussion of implications for quantum networking and multipartite quantum information processing. 
Additional technical details are provided in the appendices.

\section{Graph State and Graph Preserving Group}
\label{sec2}
In this section, we define graph-preserving Clifford operations and introduce the factorized representation used for circuit simulation and optimization. Informally, a graph-preserving operation maps products of graph-basis states to products of graph-basis states. Equivalently, it induces a permutation of the graph-basis labels associated with the target graph. This restriction is natural for distillation, where the goal is to manipulate noise within the graph-state basis while increasing the fidelity with respect to the desired graph state.

The computational value of this restriction comes from how the admissible operations are represented. Rather than storing a candidate operation as an expanded Clifford circuit or as a full stabilizer tableau on the two copies, we describe it as a fixed-size product of a small set of predetermined permitted operations, sufficient to describe a large family of possible graph-preserving gates. A gate appearing in a candidate distillation circuit can then be stored as a compact symbolic instruction, or as a pointer to a cached graph-basis update. Thus the classification of graph-preserving operations also provides the data structure used for efficient enumeration, simulation, and circuit search.

After recalling the graph-state basis and formalizing the graph-preserving condition, we identify two families of factorized operations. The first consists of homogeneous operations whose admissibility is controlled by bipartiteness; the second consists of bilocal operations supported on leaf--neighbor pairs. We refer to these families below as the \(H\) group and the \(B\) group, respectively.

\subsection{Graph State Preliminaries and Graph-Preserving Operations}

We briefly recall the definition of graph states and fix the notation used throughout the paper. We consider two copies of an $n$-qubit graph state associated with the same graph $G=(V,E)$, each copy distributed among $n$ nodes corresponding to the vertices of $G$. 
This is the natural local setting for two-qubit Clifford gates: at each node $i\in V$, one qubit from each copy is available, allowing a local two-qubit operation between corresponding vertices of the two copies. 
This viewpoint is also natural for purification, where error-detecting transformations compare information across multiple noisy copies. 
Although the gate classification is formulated in this two-copy setting, the resulting operations can be composed to build general finite-size distillation circuits using multiple input copies.

\begin{definition}[Graph state]
\label{def:graphbasis}
Let \( G = (V, E) \) be a simple undirected graph with \( n = |V| \) vertices. The corresponding $n$-qubit \emph{graph state} is defined as
\begin{equation}
\lvert G \rangle 
=\;
\prod_{(i,j) \in E} \mathrm{CZ}_{i,j}
\;\lvert + \rangle^{\otimes |V|}, 
\end{equation}
where \( \mathrm{CZ}_{i,j} \) denotes the controlled-$Z$ gate acting on qubits \( i \) and \( j \), and \( \lvert + \rangle = (\lvert 0 \rangle + \lvert 1 \rangle)/\sqrt{2} \). 
\end{definition}

The graph state \( \lvert G \rangle \) is stabilized by the generators
\begin{equation}
K_i = X_i \bigotimes_{j \in \mathrm{N}(i)} Z_j \qquad \text{for each } i \in V,
\end{equation}
where \( \mathrm{N}(i) \) denotes the neighborhood of vertex \( i \) in the graph.

\begin{definition}[Graph basis]
Any $n$-qubit state obtained from \( \lvert G \rangle \) by local Pauli operations (and hence differing from \( \lvert G \rangle \) only by stabilizer sign flips) is said to be in the \emph{graph basis}. Concretely, this basis consists of $2^n$ states of the form
\begin{equation}
\Bigl\{\, 
\Bigl(\bigotimes_{i=1}^n P_i \Bigr)
\lvert G \rangle
\;\bigm|\;
P_i \in \{I,X,Y,Z\}
\Bigr\}.    
\end{equation}
\end{definition}

For example, for the three-qubit linear graph \( 1{-}2{-}3 \), the corresponding graph state is stabilized by the generators \( X_1 Z_2 \), \( Z_1 X_2 Z_3 \), and \( Z_2 X_3 \). Any state stabilized by these operators (up to a choice of $\pm 1$ eigenvalue) lies in the same graph basis.

\begin{definition}[Graph-preserving]
\label{def:graphpreserving}
A unitary operator $U$
is called \emph{graph-preserving} if, whenever it is applied to a product of graph-basis states, the resulting output is again a product of graph-basis states. E.g., let \( \lvert G_i \rangle \) and \( \lvert G_j \rangle \) denote basis states corresponding to a fixed graph \( G \). Then \( U \) is graph-preserving if, for all such basis states,
\begin{equation}
U \Bigl(
\lvert G_i \rangle \otimes \lvert G_j \rangle
\Bigr)
=
\lvert G_k \rangle \otimes \lvert G_l \rangle,
\end{equation}
for some \( \lvert G_k \rangle \), \( \lvert G_l \rangle \) also in the graph basis.

\end{definition}

Equivalently, a graph-preserving unitary induces a permutation on the \( 2^n \times 2^n \)-dimensional basis for two multi-partite states.

Before proceeding, we introduce several graph-theoretic notions that will be used to state configuration-dependent conditions for the admissibility of graph-preserving operations. These properties depend only on the combinatorial structure of the underlying graph and will play a role in the classification developed in the following sections.

\begin{definition}[Bipartite Graph]
A graph $G = (V, E)$ is said to be \emph{bipartite} if there exists a partition of $V$ into two disjoint sets $V_1$ and $V_2$ such that every edge in $E$ connects a vertex in $V_1$ to a vertex in $V_2$.
\end{definition}

\begin{definition}[Leaf Vertex]
A vertex \( v \in V(G) \) of a graph \( G = (V, E) \) is called a \emph{leaf} if it has degree one, i.e.,
\[
\deg(v) = 1.
\]
We denote by \( \mathrm{Leaf}(G) \subseteq V(G) \) the set of all leaf vertices in \( G \), and let \( L(G) = |\mathrm{Leaf}(G)| \) denote the total number of leaf nodes.
\end{definition}
With the above definitions established, we now introduce two graph-theoretically defined classes of gates that generate the graph-preserving operations used in this work. 
These operations act as permutations of the graph basis for one or two multipartite graph states, and their availability depends on two simple graph features: bipartiteness and the presence of leaf vertices. 
They yield a compact and efficiently enumerable gate set for constructing graph-preserving distillation circuits. 
Special graph states with additional algebraic symmetries may admit further graph-preserving Clifford operations beyond this gate set; the exceptional cases are discussed in Sec.~\ref{exception}.

\begin{enumerate}
    \item The \textbf{H group}, which applies identical two-qubit gates in parallel across the system. The gates are distributed according to a pairing pattern that respects the bipartition structure of the graph.
    \item The \textbf{B group}, which consists of operations acting on only two nodes at a time, is constrained to apply only to leaf vertices.
\end{enumerate}

Together with local Pauli operations, these groups provide a natural generating set for circuits that preserve the stabilizer structure of graph states. Their algebraic formulation allows the resulting gate set to be compactly described and efficiently enumerated, which is essential for the simulation and optimization of graph-state distillation protocols. Detailed definitions and characterizations will follow in the subsequent sections, with technical proofs deferred to the Appendix.

\subsection{Homogeneous Gates (H group)}

For arbitrary graph states, the H group applies only under specific constraints. In particular, we find that the H group is available if and only if the underlying graph is bipartite.

This observation builds on earlier work by Aschauer et al.~\cite{Dur2003TwoColorable}, who developed entanglement purification protocols for bipartite graph states using only CNOT gates. In our framework, the CNOT gate is just one element of the H group, which consists of homogeneous graph-preserving operations, meaning that the same two-qubit gate is applied at every node across the system. Crucially, the entire H group is generated by the CNOT gate; therefore, if CNOT preserves the stabilizer description for bipartite graphs, so does the entire group it generates. The full set of gates is listed in Table~\ref{tab:Hgates_graph}.  

\begin{figure}[ht]
    \centering
    \includegraphics[width=0.9\linewidth]{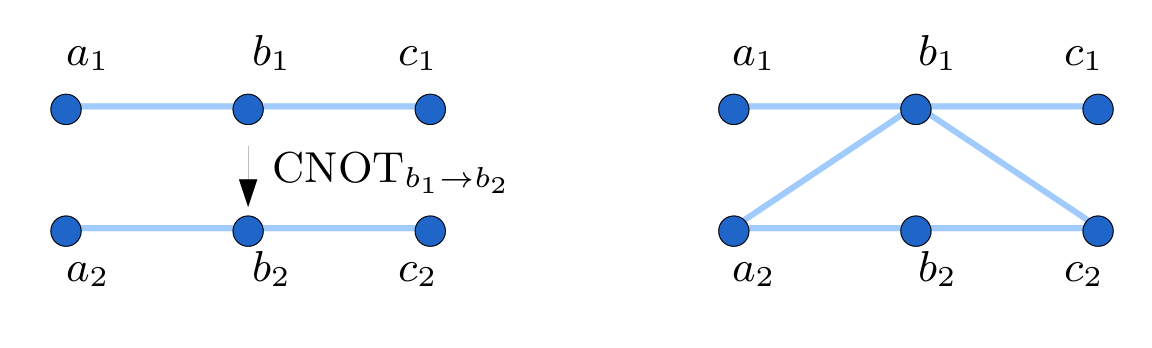}
    \caption{\textbf{Action of a CNOT gate on a graph state.}
    When a CNOT gate is applied with control qubit \(i\) and target qubit \(j\), the graph is transformed by toggling edges between the control vertex and each neighbor of the target vertex. Concretely, for every vertex ($k \in N(j)$) with ($k \neq i$), an edge \((i,k)\) is added if it is absent and removed if it is present. All other edges remain unchanged.
    }
    \label{fig:cnot_behavior}
\end{figure}

The connection to bipartiteness arises from how the CNOT gate modifies the edge connectivity of a graph state. When a CNOT gate is applied between qubits \(i\) (control) and \(j\) (target), it effectively introduces new edges between \(i\) and all neighbors of \(j\), excluding \(i\) itself~\cite{Kruszynska2006AllGraph} (see Fig.~\ref{fig:cnot_behavior}). If the graph contains an odd cycle, repeated application of such gates around the cycle introduces additional edges that accumulate rather than cancel, thereby violating the graph-preserving condition. In contrast, bipartite graphs contain no odd cycles, so the edge modifications induced by the homogeneous gates cancel pairwise across the two partitions. This cancellation mechanism is the underlying reason why the H group exists precisely for bipartite graphs.

\begin{figure}
    \centering
    \includegraphics[width=0.8\linewidth]{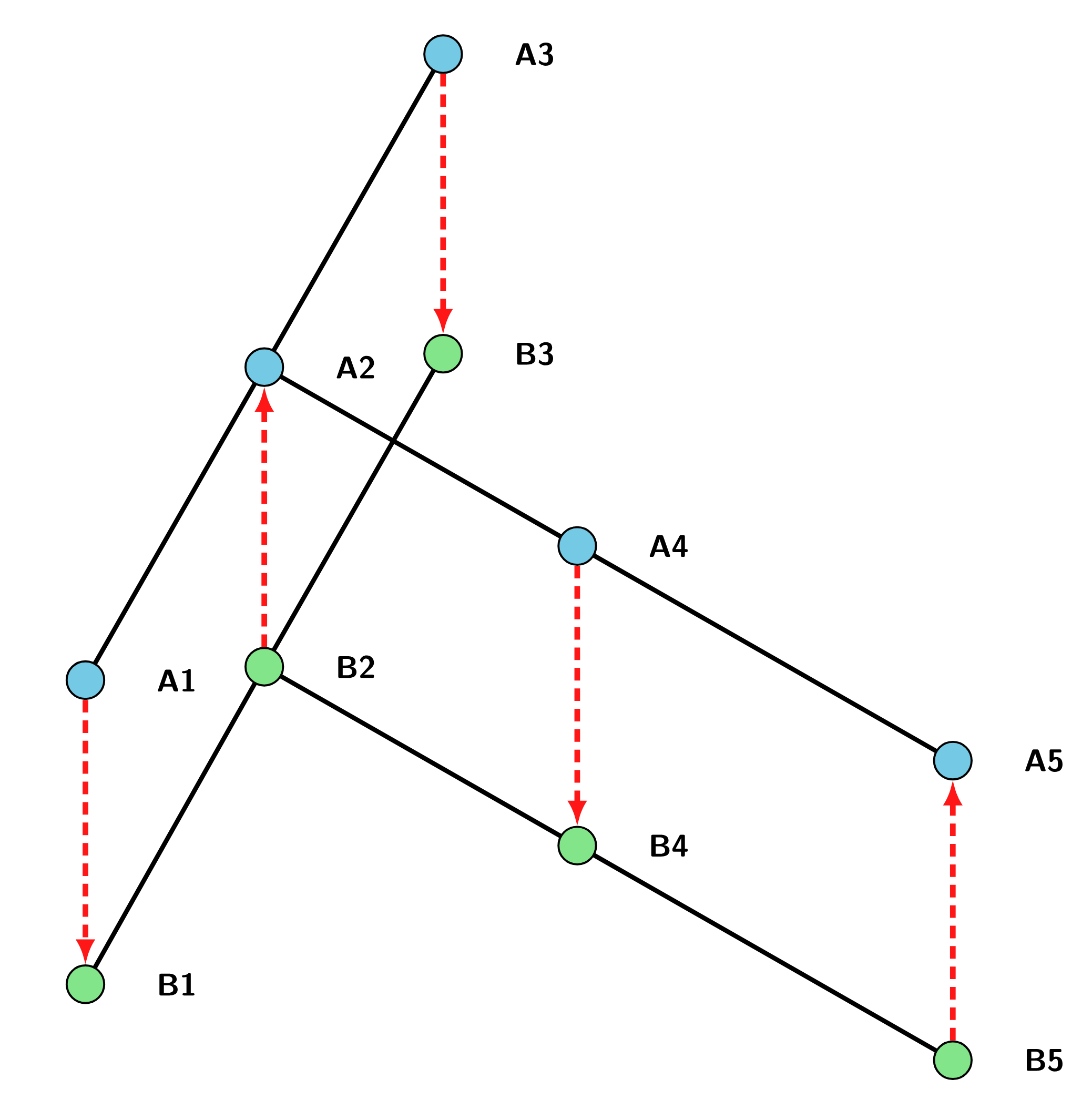}
    \caption{
    \textbf{Orientation rule for homogeneous ($H$) operations on a bipartite graph.}
    The $5$-qubit $T$-shaped graph is bipartite and can be partitioned into two disjoint vertex sets.
    To preserve the graph-state basis, homogeneous two-qubit gates must respect this bipartition:
    nodes in one partition apply the gate in one orientation, while nodes in the opposite partition apply the corresponding gate with reversed control--target direction.
    The allowed gate pairs implementing this rule are listed in Table~\ref{tab:Hgates_graph}.
    }
    \label{fig:Hgroup_5T}
\end{figure}

As illustrated in Fig.~\ref{fig:Hgroup_5T}, vertices belonging to the two partitions apply the same gate with opposite control--target orientations. This orientation rule ensures that the edge toggles induced by the local two-qubit gates cancel pairwise across the bipartite graph. Intuitively, the bipartition plays the role of a global “sign convention’’ that keeps the edge modifications introduced by each gate balanced across the graph.

The resulting group of homogeneous, bipartition-respecting operations forms a non-Abelian subgroup of the phaseless Clifford group. It has order 6 and is isomorphic to the dihedral group \(\mathbb{D}_3\).

\begin{table*}[!htbp]
\centering
\begin{tabular}{|c|c|c|c|c|c|}
\hline
XI $\to$ + XI & XI $\to$ + IX & XI $\to$ + XX & XI $\to$ + IX & XI $\to$ + XX & XI $\to$ + XI \\
IX $\to$ + IX & IX $\to$ + XI & IX $\to$ + IX & IX $\to$ + XX & IX $\to$ + XI & IX $\to$ + XX \\
ZI $\to$ + ZI & ZI $\to$ + IZ & ZI $\to$ + ZI & ZI $\to$ + ZZ & ZI $\to$ + IZ & ZI $\to$ + ZZ \\
IZ $\to$ + IZ & IZ $\to$ + ZI & IZ $\to$ + ZZ & IZ $\to$ + ZI & IZ $\to$ + ZZ & IZ $\to$ + IZ \\
\hline
\multicolumn{1}{|p{2cm}|}{\centering $Identity$} & \multicolumn{1}{p{2cm}|}{\centering $SWAP$} & \multicolumn{1}{p{2cm}|}{\centering $CNOT_{12}$} & \multicolumn{1}{p{2cm}|}{\centering $DCX_{21}$} & \multicolumn{1}{p{2cm}|}{\centering $DCX_{12}$} & \multicolumn{1}{p{2cm}|}{\centering $CNOT_{21}$} \\
\hline
\end{tabular}
\caption{
Two-qubit building-block gates generating the homogeneous ($H$) group when applied across a bipartite graph.
The identity and SWAP gates can be applied uniformly on all qubits.
For the remaining gates, the applied operation must respect the bipartition of the graph: vertices in one partition may apply any of $CNOT_{12}$, $CNOT_{21}$, $DCX_{12}$, or $DCX_{21}$, while vertices in the opposite partition must apply the corresponding complementary operation (i.e., $CNOT_{12}\leftrightarrow CNOT_{21}$ and $DCX_{12}\leftrightarrow DCX_{21}$, see Fig.~\ref{fig:Hgroup_5T}).
}
\label{tab:Hgates_graph}
\end{table*}

\subsection{Bilocal Gates (B group)}
Unlike homogeneous operations, which act in parallel across all nodes, bilocal operations act on a leaf--neighbor pair of nodes. For graph states, we find that such bilocal Clifford operations preserve the graph-state basis only under specific conditions. In particular, bilocal operations must act on \emph{leaf vertices}, i.e., vertices of degree one.

A central building block for bilocal graph-preserving operations is the two-qubit gate we refer to as \emph{XCX}, which can be viewed as a controlled-$Z$ gate in the $X$ basis. It is constructed by applying Hadamard gates to both qubits, performing a CZ, and then reversing the basis change with another pair of Hadamards:
\[
\mathrm{XCX} = (H_1 \otimes H_2)\,\mathrm{CZ}\,(H_1 \otimes H_2),
\]

\begin{figure}[ht]
    \centering
    \includegraphics[width=0.9\linewidth]{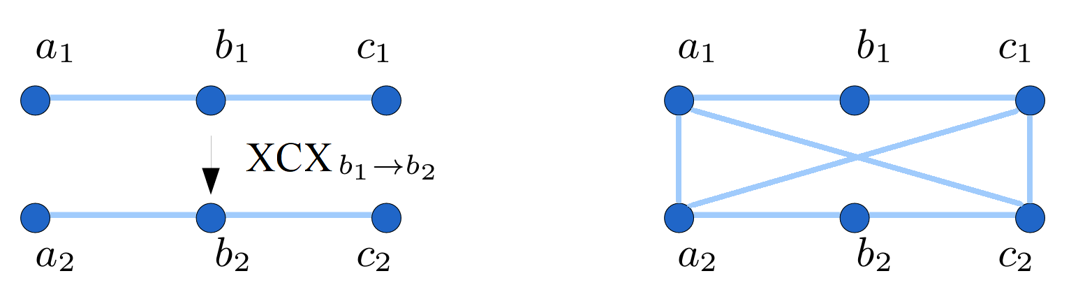}
    \caption{\textbf{Action of the XCX gate on a graph state.} 
    When an XCX gate is applied between qubits \(i\) and \(j\), it induces a complementation between the neighborhoods of the two vertices. Specifically, for every pair of vertices ($k \in N(i)$) and ($l \in N(j)$), an edge \((k,l)\) is added if it is absent and removed if it is present. All other edges remain unchanged.}
    \label{fig:xcx_behavior}
\end{figure}

From the perspective of graph structure, applying XCX between two vertices induces a complementation between their respective neighborhoods. In general, this action introduces additional edges that cannot be removed by local Clifford operations, thereby taking the state outside the graph-state basis (see Fig.~\ref{fig:xcx_behavior}).

\begin{figure}[ht]
    \centering
    \includegraphics[width=0.9\linewidth]{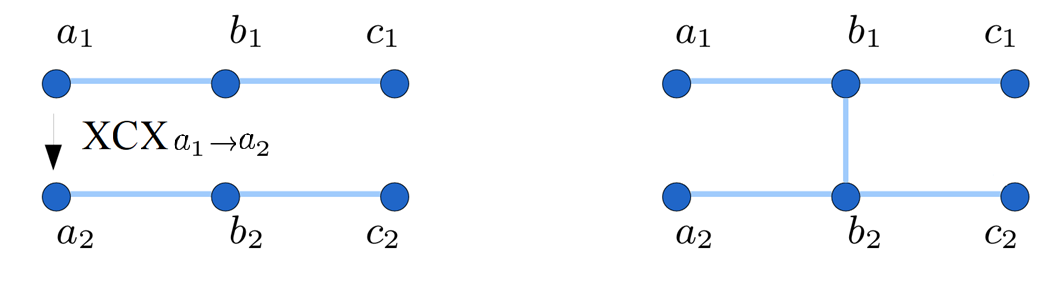}
    \caption{\textbf{Action of the XCX gate on a leaf vertex and its compensation by a CZ gate.}
    Consider an XCX gate applied between qubits \(i\) and \(j\), where \(i\) is a leaf vertex. Because \(i\) has only one neighbor, the XCX operation induces only a single edge toggle between the corresponding neighboring vertices. In the configuration shown, this creates the additional edge \((b_1,b_2)\). The extra edge can be removed by applying a CZ gate between \(b_1\) and \(b_2\), thereby restoring the original graph structure. Hence, when restricted to a leaf vertex, the effect of XCX can be exactly compensated by a CZ operation on the corresponding neighboring qubits, preserving the graph state.}
    \label{fig:xcx_leaf}
\end{figure}

When XCX acts on a leaf vertex, however, its effect simplifies dramatically. Because a leaf vertex has a single neighbor, the neighborhood complementation induced by XCX produces exactly one additional edge between the neighbors of the two acted-on qubits. This extra edge can be removed by applying a single CZ gate between those neighbors, restoring the original graph configuration (see Fig.~\ref{fig:xcx_leaf}). As a result, XCX can serve as a graph-preserving bilocal operation when applied at leaf vertices together with a compensating CZ gate.

\begin{figure}
    \centering
    \includegraphics[width=0.8\linewidth]{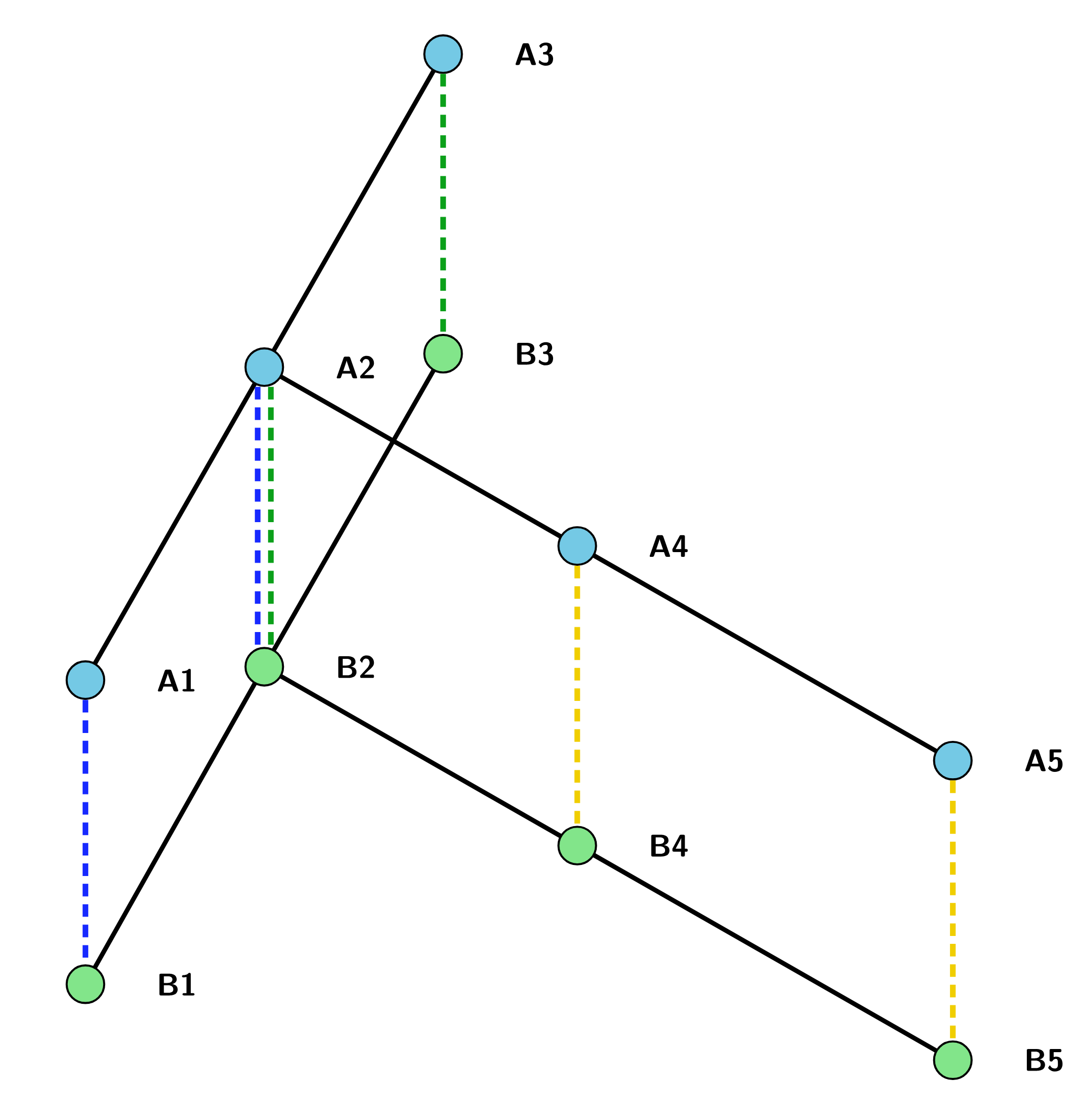}
    \caption{
    \textbf{Bilocal ($B$) operations on leaf vertices.}
    The $5$-qubit $T$-shaped graph state contains three leaf vertices. 
    Each leaf together with its unique neighbor forms a pair on which a bilocal gate can act across the two copies of the state. 
    Consequently, up to three $B$-group operations may be applied independently, one at each leaf-neighbor pair. 
    For each such pair, the gate acting on the leaf qubit is chosen from the set listed in Table~\ref{tab:Bgates_leaf}, while the corresponding operation applied to the neighbor qubit follows the mapping given in Table~\ref{tab:Bgates_neighbor}. 
    Because each leaf connects to the rest of the graph through a single edge, these bilocal operations remain confined to the corresponding leaf-neighbor pair and do not propagate to other parts of the graph.
    }
    \label{fig:Bgroup_5T}
\end{figure}

This mechanism naturally leads to the general form of bilocal graph-preserving operations. For every leaf vertex and its unique neighbor, a pair of operations acts on the corresponding qubits of the two graph-state copies: one gate acts on the leaf qubit across the two copies, and another gate acts on the neighboring qubit across the two copies. As illustrated in Fig.~\ref{fig:Bgroup_5T}, graphs with multiple leaves therefore allow several such bilocal operations to be applied independently, one for each leaf-neighbor pair. For each pair, the gate acting on the leaf qubit may be freely chosen from the set listed in Table~\ref{tab:Bgates_leaf}. Once this choice is made, the gate applied on the neighboring qubit is fixed by the corresponding entry in Table~\ref{tab:Bgates_neighbor}. The two tables therefore specify paired operations on the leaf and its neighbor that together preserve the graph-state basis.

The same pairing principle applies to single-qubit phase operations. When a gate is applied on a leaf vertex, a corresponding gate must be applied on its neighboring vertex in order to preserve the graph-state basis. For example, applying an $SHS$ gate on a leaf requires an $S$ gate on the neighboring vertex. More generally, each admissible gate acting on a leaf vertex uniquely determines the gate that must act on its neighbor. The allowed choices for the leaf qubit are listed in Table~\ref{tab:Bgates_leaf}, while the corresponding gates on the neighboring qubit are given in Table~\ref{tab:Bgates_neighbor}. 
These two tables therefore specify the complete set of bilocal gate pairs that preserve the graph-state basis. For instance, the $XCX$ operation acting on a leaf corresponds to a $CZ$ gate acting on the neighboring vertex, reflecting the same basis transformation between the two sides of the leaf–neighbor pair.

Fundamentally, the bilocal gate pairs introduced above are graph-preserving by construction: their joint action maps graph-basis states to graph-basis states.  
The restriction to leaf vertices ensures that any intermediate modification induced by a bilocal operation remains confined to a single leaf–neighbor pair and can therefore be exactly compensated by a local adjustment on the neighboring vertex. Algebraically, the resulting set of bilocal graph-preserving operations forms an Abelian group isomorphic to $\mathbb{Z}_2 \otimes \mathbb{Z}_2 \otimes \mathbb{Z}_2$, corresponding to three independent binary choices associated with the admissible bilocal gate components.

\begin{table*}[!htbp]
\centering
\begin{tabular}{|c|c|c|c|}
\hline
XI $\to$ + XI & XI $\to$ + XI & XI $\to$ + XI & XI $\to$ + XI \\
IX $\to$ + IX & IX $\to$ + IX & IX $\to$ + IX & IX $\to$ + IX \\
ZI $\to$ + ZI & ZI $\to$ + ZI & ZI $\to$ + YI & ZI $\to$ + YI\\
IZ $\to$ + IZ & IZ $\to$ + IY & IZ $\to$ + IZ & IZ $\to$ + IY\\
\hline
\multicolumn{1}{|p{3.5cm}|}{\centering $Identity$} & \multicolumn{1}{p{3.5cm}|}{\centering $I \otimes SHS$} & \multicolumn{1}{p{3.5cm}|}{\centering $SHS \otimes I$} & \multicolumn{1}{p{3.5cm}|}{\centering $SHS \otimes SHS$}

\end{tabular}
\begin{tabular}{|c|c|c|c|}
\hline
XI $\to$ + XI & XI $\to$ + XI & XI $\to$ + XI & XI $\to$ + XI \\
IX $\to$ + IX & IX $\to$ + IX & IX $\to$ + IX & IX $\to$ + IX \\
ZI $\to$ + ZX & ZI $\to$ + ZX & ZI $\to$ + YX & ZI $\to$ + YX \\
IZ $\to$ + XZ & IZ $\to$ + XY & IZ $\to$ + XZ & IZ $\to$ + XY \\
\hline
\multicolumn{1}{|p{3.5cm}|}{\centering $XCX$} & \multicolumn{1}{p{3.5cm}|}{\centering $XCX \cdot I \otimes SHS$} & \multicolumn{1}{p{3.5cm}|}{\centering $XCX \cdot SHS \otimes I$} & \multicolumn{1}{p{3.5cm}|}{\centering $XCX \cdot SHS \otimes SHS$} \\
\hline
\end{tabular}
\caption{
Allowed gates acting on the \emph{leaf vertex} of a leaf-neighbor pair that generate the bilocal ($B$) group.
Each cell specifies one possible gate applied on the two leaf qubits at the given location (one from each of the two distributed graph states).
A leaf vertex may freely choose any one of the eight gates listed here.
Once a gate is chosen, the qubit at the neighboring vertex must apply the corresponding gate from Table~\ref{tab:Bgates_neighbor}.
}
\label{tab:Bgates_leaf}
\end{table*}

\begin{table*}[t]
\centering
\begin{tabular}{|c|c|c|c|}
\hline
XI $\to$ + XI & XI $\to$ + XI & XI $\to$ + YI & XI $\to$ + YI \\
IX $\to$ + IX & IX $\to$ + IY & IX $\to$ + IX & IX $\to$ + IY \\
ZI $\to$ + ZI & ZI $\to$ + ZI & ZI $\to$ + ZI & ZI $\to$ + ZI\\
IZ $\to$ + IZ & IZ $\to$ + IZ & IZ $\to$ + IZ & IZ $\to$ + IZ\\
\hline
\multicolumn{1}{|p{3cm}|}{\centering $Identity$} & \multicolumn{1}{p{3cm}|}{\centering $I \otimes S$} & \multicolumn{1}{p{3cm}|}{\centering $S \otimes I$} & \multicolumn{1}{p{3cm}|}{\centering $S \otimes S$}

\end{tabular}
\begin{tabular}{|c|c|c|c|}
\hline
XI $\to$ + XZ & XI $\to$ + XZ & XI $\to$ + YZ & XI $\to$ + YZ \\
IX $\to$ + ZX & IX $\to$ + ZY & IX $\to$ + ZX & IX $\to$ + ZY \\
ZI $\to$ + ZI & ZI $\to$ + ZI & ZI $\to$ + ZI & ZI $\to$ + ZI \\
IZ $\to$ + IZ & IZ $\to$ + IZ & IZ $\to$ + IZ & IZ $\to$ + IZ \\
\hline
\multicolumn{1}{|p{3cm}|}{\centering $CZ$} & \multicolumn{1}{p{3cm}|}{\centering $CZ \cdot I \otimes S$} & \multicolumn{1}{p{3cm}|}{\centering $CZ \cdot S \otimes I$} & \multicolumn{1}{p{3cm}|}{\centering $CZ \cdot S \otimes S$} \\
\hline
\end{tabular}
\caption{
Corresponding gates applied on the \emph{neighbor vertex} for the bilocal ($B$) operations listed in Table~\ref{tab:Bgates_leaf}.
Unlike the leaf vertex, the neighbor vertex cannot choose its gate freely: for each cell in Table~\ref{tab:Bgates_leaf}, the gate applied on the neighbor qubit must be the gate shown in the corresponding cell of this table.
The two tables therefore define paired operations on a leaf-neighbor pair.
The gates appearing here can be viewed as basis-adjusted versions of those in Table~\ref{tab:Bgates_leaf}; for example, the gate $XCX$ in the leaf table corresponds to a $CZ$ gate on the neighbor vertex, and $SHS$ corresponds to $S$ under the same basis adjustment.
}
\label{tab:Bgates_neighbor}
\end{table*}

\begin{table*}[t]
\centering
\renewcommand{\arraystretch}{1.5}
\setlength{\tabcolsep}{8pt}
\begin{tabular}{|c|c|c|c|}
\hline
\multicolumn{2}{|c|}{\textbf{Graph Configuration}} 
& \multicolumn{2}{c|}{\textbf{H group}} \\
\cline{3-4}
\multicolumn{2}{|c|}{} 
& \textbf{Bipartite} 
& \textbf{Non-bipartite} \\
\hline
\multirow{2}{*}{\textbf{B group}} 
& \textbf{Leaf vertices present} 
& \includegraphics[width=2cm]{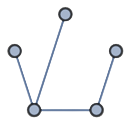}
& \includegraphics[width=2cm]{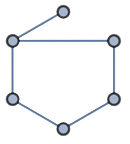} \\
\cline{2-4}
& \textbf{No leaf vertices} 
& \includegraphics[width=2cm]{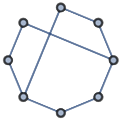}
& \includegraphics[width=2cm]{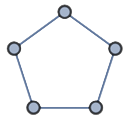} \\
\hline
\end{tabular}
\caption{\textbf{Gate admissibility determined by bipartiteness and leaf structure.} Illustration of the admissibility of graph-preserving operations as determined by two graph-theoretic properties: bipartiteness and the presence of leaf vertices.
The columns distinguish whether the graph is bipartite, which determines the availability of homogeneous ($H$) operations.
The rows distinguish whether leaf vertices are present, which determines the availability of bilocal ($B$) operations.
Top-left: bipartite graph with leaves, both $H$ and $B$ groups are available.
Top-right: non-bipartite graph with leaves, only $B$ group operations are available.
Bottom-left: bipartite graph without leaves, only $H$ group operations are available.
Bottom-right: non-bipartite graph without leaves, neither $H$ nor $B$ operations are available.
}
\label{tab:group_admissibility}
\end{table*}

\subsection{Factorized Graph-Preserving Gates}

The previous subsections characterize the two fundamental classes of graph-preserving operations: homogeneous ($H$) operations determined by bipartiteness and bilocal ($B$) operations determined by leaf vertices.
We now combine these ingredients into a unified form for graph-preserving Clifford gates.

Consider two instances of an $n$-qubit state from the graph-basis associated with a graph $G=(V,E)$.
A generic graph-preserving operation is a multilocal Clifford gate built from three components:
\begin{itemize}
    \item local Pauli operations acting independently on the qubits of the two states,
    \item paired bilocal $B$-group operations acting on each leaf vertex and its unique neighbor across the two states (see Fig.~\ref{fig:Bgroup_5T}),
    \item homogeneous $H$-group operations acting across all vertices when the graph is bipartite (see Fig.~\ref{fig:Hgroup_5T}).
\end{itemize}

\begin{figure*}[t]
    \centering

    \begin{minipage}[t]{0.32\textwidth}
        \centering
        \includegraphics[width=\linewidth]{H_group_T.png}
        \\[-1.0\baselineskip]
        \textbf{(a)} 
        \includegraphics[width=\linewidth]{B_group_T.png}
        \\[-1.0\baselineskip]
        \textbf{(b)} 
    \end{minipage}%
    \hfill
    \begin{minipage}{0.65\textwidth}
        \centering
        
        \raisebox{-\height}{\includegraphics[width=\linewidth]{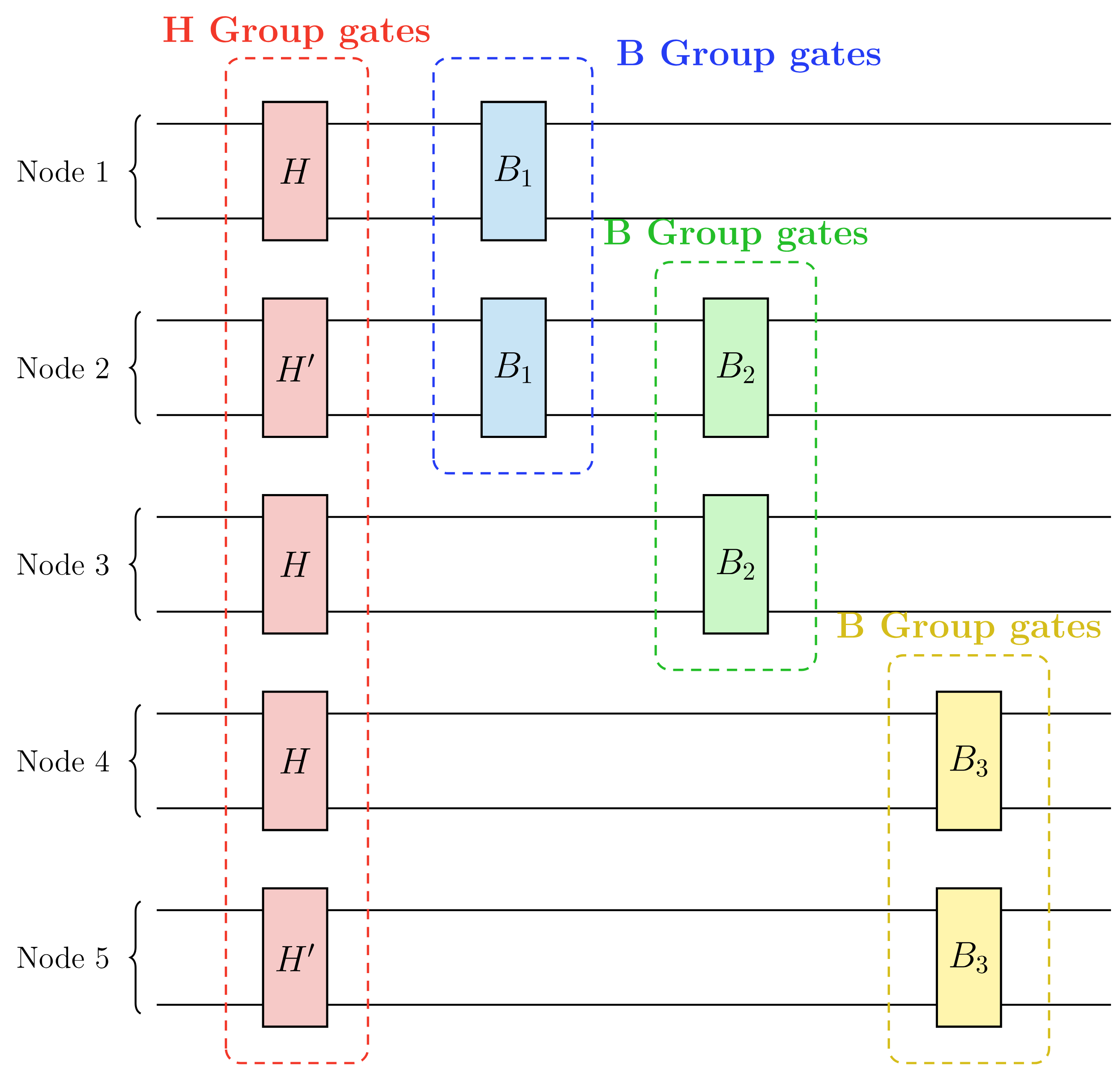}}
        \textbf{(c)} 
    \end{minipage}

    \vspace{0.8em}

    \caption{\textbf{Example of a graph-preserving circuit acting on two copies of a $5$-qubit $T$-shaped graph state.}
    Two identical graph states are shown, each distributed across five nodes.
    Qubits $(1,3,5,7,9)$ correspond to the first graph state, and $(2,4,6,8,10)$ to the second state. Each node therefore holds two qubits, one from each state.
    \textbf{(a)}~\emph{Homogeneous operations (H group):}
    The underlying graph is bipartite, and the nodes are partitioned into two disjoint sets. To preserve the graph basis, homogeneous two-qubit gates must respect this bipartition: nodes in one partition apply the gate in one orientation, while nodes in the other partition apply the same gate with reversed control-target direction. This orientation constraint is explained in Fig.~\ref{fig:cnot_behavior}. See Table~\ref{tab:Hgates_graph} for the allowed gate set.
    \textbf{(b)}~\emph{Bilocal operations (B group):}
    Each leaf vertex may independently host a bilocal gate. At each such location, one may choose any element of the $B$ group (see Table~\ref{tab:Bgates_leaf}). Together with local Pauli operations (not shown), these gates generate the factorized graph-preserving circuits for this graph configuration.
    \textbf{(c)}~\emph{Example circuit.}
    The circuit implements a graph-preserving transformation on two copies of the state.
    Gates labeled $H$ and $H'$ apply the homogeneous operation in opposite orientations
    according to panel~(a), while nodes labeled $B$ host bilocal operations according to panel~(b).
    At the level of simulation and optimization, each labeled block is stored as a compact indexed circuit instruction whose graph-basis action is retrieved from a precomputed cache, rather than as an expanded Clifford tableau. Local Pauli corrections are omitted for clarity.
    }

    \label{fig:circuits}

\end{figure*}

More explicitly, such an operation may be written in the form
\begin{equation}
\label{factor}
g =
\Bigl(\prod_{j\in V} p_j^{(1)} p_j^{(2)} \Bigr)
\Bigl(\prod_{v \in \mathrm{Leaf}(G)} B_{v,\nu(v)} \Bigr)
H_{V_A,V_B}^{\delta},
\end{equation}

where $p_j^{(1)}$ and $p_j^{(2)}$ are local Pauli operators acting on qubit $j$ in the first and second copy, respectively; $\nu(v)$ denotes the unique neighbor of the leaf vertex $v$; $B_{v,\nu(v)}$ denotes the pair of bilocal operations associated with the leaf--neighbor pair $(v,\nu(v))$, consisting of one gate acting on the $v$ qubits of each of the two graph states (see Table~\ref{tab:Bgates_leaf}), and the corresponding second gate acting on the $\nu(v)$ qubits of each of the two graph states (see Table~\ref{tab:Bgates_neighbor}); and $H_{V_A,V_B}$ denotes the homogeneous operation associated with a bipartition $V=V_A\sqcup V_B$, such that the vertices in $V_A$ and $V_B$ experience gates as given in the gate pairings in Table~\ref{tab:Hgates_graph}. The exponent $\delta\in\{0,1\}$ indicates whether $G$ is bipartite, so that the homogeneous contribution is present only in that case. See Fig.~\ref{fig:circuits}.

We call gates of the form in Eq.~\ref{factor} \emph{factorized graph-preserving gates}. The terminology is meant both algebraically and computationally. Algebraically, a gate is assembled from independent factors (dependent on the particular graph): local Pauli-frame updates, leaf-supported \(B\)-group components, and, when available, a bipartition-supported \(H\)-group component. Computationally, this means that the operation need not be stored or manipulated as an expanded Clifford circuit on the \(2n\) physical qubits of the two copies, but as a much more compact fixed-size product of predefined "basic" operations.

For a fixed graph representative, we precompute the graph-basis action of the elementary \(H\) and \(B\) factors. A gate appearing in an optimized circuit is then represented by a compact descriptor, or equivalently by a pointer into this graph-specific cache. During simulation, the gate is evaluated by retrieving and composing cached label maps, rather than reconstructing the Clifford action from stabilizer generators. The precomputed cache remains small because it is built factorwise: one stores the six possible \(H\)-group actions when \(\delta=1\), together with the eight possible \(B\)-group actions for each admissible leaf--neighbor pair, instead of storing all expanded \(2n\)-qubit Clifford realizations or all candidate circuits.

This factorization is the main reason the gate set is useful for optimization. The set is restricted relative to the full transversal Clifford search space, and exceptional graph states may admit additional Clifford symmetries beyond the \(H/B\) construction. Nevertheless, the factorized set is large enough to produce nontrivial distillation circuits while remaining small enough to enumerate, cache, and repeatedly apply inside an automated search.

Ignoring local Pauli corrections, the number of formal products generated by this generic construction is
\begin{equation}
    |G_n| = 6^\delta \cdot 8^t ,
\end{equation}
where \(t\) is the number of leaf vertices and \(\delta=1\) only when the graph is bipartite. This count should be understood as the size of the generic factorized gate family, not as a requirement to store every element as an independent expanded Clifford operation. In implementation, the factors are cached separately and composed by reference during circuit evaluation.

To reiterate, by precomputing $6+8$ gates, we are able to simulate all $6\cdot 8^t$ graph-preserving gates much faster than the Clifford formalism, enabling the circuit optimization steps discussed in later paragraphs.

\section{Graph Preserving Group on Minimum Edge Representatives (MERs)}
\label{sec3}
In the previous section, we introduced the homogeneous ($H$) and bilocal ($B$) groups as two families of graph-preserving operations defined on a fixed graph configuration. These operations already allow for nontrivial transformations within the graph-state basis, but their applicability depends explicitly on structural properties of the underlying graph, such as bipartiteness and the presence of leaf vertices. As a result, different graph configurations representing locally equivalent graph states may appear to admit different sets of graph-preserving operations.

This observation motivates a refinement of the classification: although locally equivalent graph states describe the same physical entanglement, their explicit graph configurations can differ substantially, and with them the apparent availability of $H$ and $B$ group operations. To address this mismatch, we introduce local complementation (LC) and the associated LC orbit, which captures all graph configurations related by local Clifford equivalence. The LC equivalence is well studied in the literature on graph states, and the reader can consult~\cite{VanDenNest2004LC,VanDenNest2005LUvsLC,Adcock2020} for further background. Within each LC orbit, we further identify a Minimum Edge Representative (MER), which serves as a canonical reference among equivalent representations. By analyzing graph-preserving gates at the level of LC orbits via their MERs, we obtain a unified and consistent description of the admissible gate sets used in this work.

This section is organized as follows. In Sec.~\ref{lc_mer_def} we review local complementation and introduce the MER as a canonical representative within each LC orbit. This part primarily serves to fix notation and recall relevant background, while formalizing the MER construction used throughout the paper. In Sec.~\ref{complexity} we discuss computational aspects, including complexity considerations and a database-based workflow that enables practical use of MERs in circuit design. We then return to the structure of graph-preserving operations in Sec.~\ref{complete_decomposition}, where we show how the MER provides a consistent reference for determining the admissible $H$ and $B$ group operations across an entire LC orbit, and derive a general gate-counting formula, including connections to known special cases such as GHZ states.

\subsection{Local Complementation and Minimum Edge Representatives (MERs)}
\label{lc_mer_def}
\begin{figure}
    \centering
    \includegraphics[width=0.95\linewidth]{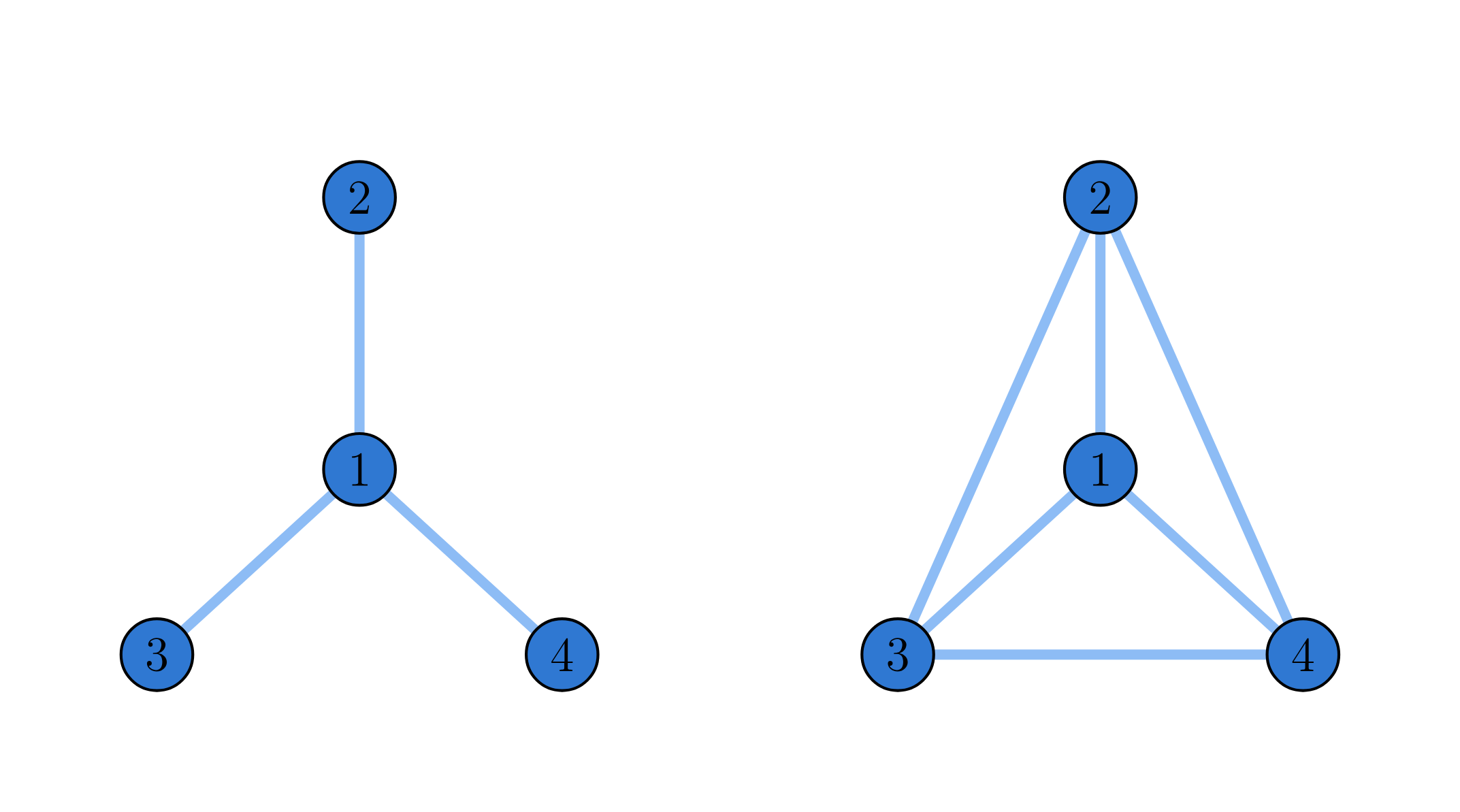}
    \caption{\textbf{Local Complementation at a Vertex.}~Applying local complementation about vertex 1 complements all adjacencies within its neighborhood $\{2, 3, 4\}$. \textbf{Left:} The induced subgraph is edgeless. \textbf{Right:} The operation converts it into a clique. As the operation is self-inverse, applying it again restores the original graph. Both graphs are locally equivalent to a 4-qubit GHZ state.}
    \label{fig:lc_operation}
\end{figure}

Local complementation (LC) is a graph operation that transforms a graph $G$ by complementing the subgraph induced by the neighborhood of a chosen vertex~(Fig.~\ref{fig:lc_operation}). At the level of graph states, LC corresponds to conjugation by an appropriate \emph{local Clifford} unitary, mapping $|G\rangle$ to another stabilizer state $|G'\rangle$ that is locally Clifford equivalent to $|G\rangle$ (up to global phase)~\cite{Hein2004Multiparty,VanDenNest2004LC}. This correspondence allows us to consider the \emph{LC orbit} of a graph: the set of graph configurations reachable from $G$ via sequences of local complementations, all representing locally equivalent graph states while potentially exhibiting different graph-theoretic properties~\cite{Adcock2020}.

\begin{definition}[Local Complementation (LC)]
Given a graph $G = (V, E)$ and a vertex $v \in V$, the \emph{local complementation} at $v$, denoted $G^v$, is defined as the graph obtained by replacing the subgraph induced by the neighborhood $N_v$ with its complement. That is, for every pair $u, w \in N_v$, if $\{u, w\} \in E$ then it is removed, and if $\{u, w\} \notin E$ then it is added. Note that $(G^v)^v = G$.
\end{definition}

In the stabilizer formalism, local complementation corresponds to applying the square root of the stabilizer generator associated with vertex $i$, which takes the form:
\begin{equation}
\sqrt{S_i} = \sqrt{X_i} \bigotimes_{j \in n_i} \sqrt{Z_j}
\end{equation}
where
\begin{equation}
    \sqrt{Z} = \begin{bmatrix} 1 & 0 \\ 0 & i \end{bmatrix} = S, \qquad
    \sqrt{X} = \frac{1}{2} \begin{bmatrix} 1+i & 1-i \\ 1-i & 1+i \end{bmatrix} = H S H
\end{equation}

By repeatedly applying LC operations, we generate a set of graphs that are all locally Clifford equivalent; this set is called the \emph{LC orbit}. The LC orbit collects all graph configurations that represent the same graph state up to local Clifford transformations (i.e.\ the orbit includes many distinct graphs, but the corresponding quantum graph states are equivalent under local operations) and provides the natural basis for equivalence classification and the identification of canonical representatives.

\begin{definition}[LC Orbit]
Given a graph \( G \), its \emph{LC orbit} is the set of all graphs obtained from $G$ by applying a finite sequence of LC operations:
\begin{align}
    \mathcal{O}_{\mathrm{LC}}(G) &= \{ G' \;|\; G' = \tau_{i_k} \circ \cdots \circ \tau_{i_1}(G) \nonumber \\
    &\quad \text{for some sequence of vertices } i_1, \dots, i_k \}.
\end{align}

Two graphs \( G \) and \( G' \) are \emph{LC-equivalent} if they lie in the same LC orbit.
\end{definition}

It is important to distinguish between two types of LC orbits:
\begin{enumerate}
    \item The \emph{labeled LC orbit} of a graph $G$, defined as the set of graphs obtained from $G$ by sequences of local complementations, with vertex labels fixed.

    \item The \emph{unlabeled (or isomorphic) LC orbit}, defined as the equivalence class of the labeled LC orbit under graph isomorphism. That is, two graphs are identified if they are isomorphic as unlabeled graphs.
\end{enumerate}

Graphs that are identical up to a relabeling of their vertices are said to be \emph{isomorphic}. Distinguishing such isomorphic graphs can be important during analysis in settings where qubit labels are fixed by hardware constraints or network architecture and cannot be freely reassigned.

In what follows, all discussions and classifications will be based on the \emph{isomorphic} LC orbit, unless otherwise specified. This allows us to treat graphs up to isomorphism and focus on structural properties independent of vertex labeling. When needed, one can recover specific labeled representatives by considering explicit vertex permutations in addition to LC operations.

Within an LC orbit, different graphs may exhibit vastly different edge structures, even though they represent locally equivalent entangled states. This variation motivates the identification of a canonical representative--one that minimizes a well-defined structural quantity. In this work, we consider the number of edges as such a quantity and define the \emph{Minimum Edge Representative} accordingly.

\begin{definition}[Minimum Edge Representative (MER)]
Given a graph \( G \), its \emph{Minimum Edge Representative} within the LC orbit is defined as
\[
\mathrm{MER}(G) = \arg\min_{G' \in \mathcal{O}_{\mathrm{LC}}(G)} |E(G')|,
\]
where the minimization is over graphs up to isomorphism, and \( |E(G')| \) denotes the number of edges in \( G' \).
\end{definition}

The MER construction provides a compact proxy for an LC equivalence class by identifying the sparsest graph configurations within the orbit. Since each LC orbit on a fixed number of vertices is finite, \( \mathrm{MER}(G) \) is nonempty for every graph \(G\). In general, the minimum edge count need not be achieved by a unique graph configuration: the set \( \mathrm{MER}(G) \) may contain multiple non-isomorphic graphs, and each such graph corresponds to many labeled representatives under vertex relabelings. When a unique representative is required, we select a canonical element from \( \mathrm{MER}(G) \) using the canonical indexing scheme introduced in Ref.~\cite{Cabello2011}.

\subsection{Computational Considerations and a Database Viewpoint}
\label{complexity}
Identifying a minimum-edge representative (MER) of an LC orbit is an orbit-level optimization problem: one must minimize the edge count over all graph representatives in the orbit. 
To our knowledge, there is no known efficient algorithm that directly solves this minimization problem for arbitrary graphs. 
A direct way to certify that a given representative is minimal is to compare its edge count against all representatives in the LC orbit, whose size can grow rapidly with the number of vertices. 
This difficulty is consistent with the broader computational complexity of related graph-state transformation and orbit-enumeration problems. 
In particular, deciding whether a target graph state can be generated from an input graph state using LC operations, local Pauli measurements, and classical communication is NP-complete for both labelled and unlabelled graphs~\cite{Dahlberg2018TransformGraphStates,dahlberg2019vertexminor,Dahlberg2018TransformingGraphStates}, and counting single-qubit LC-equivalent graph states is \#P-complete~\cite{Dahlberg2020CountingCliffordEquivalent}. 
At the same time, deciding whether two specified graph states are LC-equivalent, and finding a corresponding LC unitary when one exists, can be done in polynomial time using known algorithms based on isotropic-system techniques~\cite{bouchet1991efficient,bouchet1993recognizing,LCequv}. 

This separation motivates a practical workflow aligned with our framework. The MERs of LC orbits can be computed once in an offline precomputation stage for the graph sizes of interest (e.g.\ practically $n<20$), together with associated metadata such as admissible graph-preserving gates or optimized distillation templates. Subsequently, given an arbitrary input graph state $|G\rangle$, one can rapidly identify the corresponding LC orbit by comparing $G$ against entries in this database using LC-equivalence checks. Once the matching MER is identified, the corresponding pre-optimized distillation circuit can be instantiated immediately, with only minor adjustments for hardware-dependent parameters such as noise models or register constraints. In this way, the computationally demanding part of the problem is shifted to a one-time preprocessing step, while the online identification and circuit deployment remain fast in practice.

While this approach is limited to relatively small system sizes, this regime is precisely the one most relevant for upcoming modular architectures. In particular, in measurement-based quantum computation (MBQC) and quantum networking protocols, large graph states are often assembled probabilistically from smaller resource states. The probabilistic nature of these processes is what makes purification techniques (like the ones we are developing here) particularly applicable. In MBQC settings, the availability of highly optimized purification circuits for these small building-block graph states can significantly improve overall performance by increasing the success probability and effective fidelity of the construction process. The framework developed here provides a systematic way to generate and reuse such optimized building-block circuits.

\subsection{MER-Based Decomposition and Gate Counting}

\label{complete_decomposition}
The applicability of the $H$ and $B$ groups depends on simple graph-theoretic features of the underlying graph. The $H$ group is available when the graph is bipartite, while each $B$ group operation is associated with a leaf vertex and its unique neighbor. Thus, for any fixed graph configuration, the admissible graph-preserving gates can be determined by checking bipartiteness and counting leaf vertices.

\begin{figure*}[t]
    \centering
    
    \hspace{0.1\linewidth}
    \includegraphics[width=0.8\linewidth]{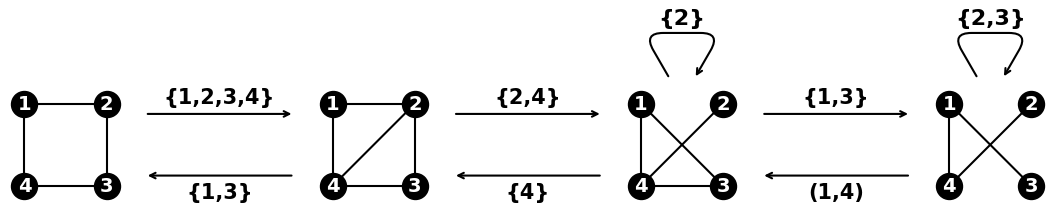}
    
    \vspace{4pt}
    
    {\raggedright
    \renewcommand{\arraystretch}{1.4}
    \begin{tabular}{ccccc}
    \textbf{Bipartite} 
    & \hspace{0.5cm}Yes 
    & \hspace{3.5cm}No 
    & \hspace{3.5cm}No 
    & \hspace{3.5cm}Yes \\
    
    \textbf{Leaf Count} 
    & \hspace{0.5cm}0 
    & \hspace{3.5cm}0 
    & \hspace{3.5cm}1 
    & \hspace{3.5cm}2 \\
    
    \textbf{Group} 
    & \hspace{0.5cm}H 
    & \hspace{3.5cm}N/A 
    & \hspace{3.5cm}B 
    & \hspace{3.5cm}H\&B
    \end{tabular}
    }
    
    \caption{
    \textbf{Variation of admissible graph-preserving operations within an LC orbit.}
    The figure shows an isomorphic LC orbit of a four-vertex graph. Although all graphs in the orbit correspond to locally equivalent graph states, their graph representations differ.
    In particular, bipartiteness and the number of leaf vertices vary across the orbit.
    The table lists these properties for each representative and indicates the corresponding availability of $H$ and $B$ group operations. This example illustrates that the admissible gate set depends on the chosen graph representation within the LC orbit, motivating the use of the minimum-edge representative (MER) as a canonical reference for determining the largest accessible set of graph-preserving operations built from instances of the $H$ and $B$ groups.
    }
    \label{fig:lc-orbit-table}
\end{figure*}

However, this description is not yet sufficient for an LC-orbit-level framework. Graph states in the same LC orbit are locally Clifford equivalent and therefore represent the same entanglement resource, but their graph configurations may have different bipartiteness and leaf structure. Figure~\ref{fig:lc-orbit-table} illustrates this explicitly: several LC-equivalent representatives of the same four-vertex graph state admit different apparent subsets of $H$ and $B$ operations. If the admissible gate set were assigned directly from an arbitrary representative, the resulting description would depend on a graph presentation rather than on the LC equivalence class.

To remove this ambiguity, we assign the generic graph-preserving gate set by evaluating the relevant graph properties on a minimum-edge representative (MER). The MER is not required for graph-preservation itself: any graph representative with a known bipartition or leaf structure gives a valid graph-preserving subgroup. Rather, the MER provides a canonical orbit-level reference on which these properties are usually easiest to expose. In particular, sparsifying the graph often reveals degree-one vertices that are not visible in denser representatives, thereby making more $B$ operations available; see Table~\ref{tab:lc_orbit_leaves}. Similarly, when a bipartite representative exists in the orbit, the MER viewpoint gives a natural place to record the corresponding availability of the $H$ group; numerical evidence and the orbit databases used in this work are discussed in Appendix~\ref{app:LC-MER}.

The MER therefore serves as the canonical reference for assigning graph-preserving gates within an LC orbit: admissibility of the $H$ and $B$ groups is determined once from MER properties, yielding a gate set that is invariant under vertex relabeling and local Clifford transformations within the orbit. With this convention, modulo local Pauli correction, the size of the graph-preserving group associated with an $n$-vertex graph state can be written as
\begin{equation}
    \begin{aligned}
    |G_{\mathrm{MER}}| &= |H|^{\delta} \cdot |B|^{t} = 6^{\delta} \cdot 8^{t},\\
    \text{where} \quad
    \delta &= 
    \begin{cases}
        1, & \text{if the MER is bipartite}, \\
        0, & \text{otherwise},
    \end{cases} \\
    t &= \text{the number of leaf vertices in the MER}.
    \end{aligned}
    \label{eq:generic-count}
\end{equation}

Importantly, this expression is not restricted to minimum-edge representatives. For any given graph state, one may substitute its own bipartiteness and leaf count directly into the above formula to obtain a valid graph-preserving subgroup for that particular graph configuration without invoking local complementation.
The role of the MER is instead to choose a consistent and typically more favorable representative for organizing the admissible gate set across the whole LC orbit.

Operationally, once the admissible gate is specified on the MER, it can be used for any locally equivalent graph by conjugating with the local Clifford maps that relate the graph to the MER. Thus the MER functions as a common coordinate system for assigning graph-preserving operations to an LC orbit, while preserving the freedom to instantiate the resulting circuit on the representative most suitable for a given hardware layout or noise model.

\paragraph*{Consistency with the GHZ Case.}
The formula~\eqref{eq:generic-count} naturally reproduces the known structure of GHZ-preserving operations~\cite{GHZPreserving}. An $n$-qubit GHZ state corresponds to a star graph, which is bipartite ($\delta = 1$) and has $t=n-1$ leaf vertices, giving
\begin{equation}
    |G_n^{\mathrm{GHZ}}| = 6 \cdot 8^{n-1},
\end{equation}
in exact agreement with the homogeneous--bilocal decomposition of that work. The GHZ case is therefore recovered as a special instance of the present graph-state framework.

\paragraph*{Scope of the generic gate set.}
\label{exception}
Equation~\eqref{eq:generic-count} describes the \emph{generic} graph-preserving gate sets arising from bipartiteness and leaf-based bilocal operations. It does not attempt to characterize rare exceptional graphs that admit additional Clifford symmetries beyond those generated by the $H$ and $B$ groups.

A representative example is the $8$-vertex cube graph $Q_3$. This graph is bipartite ($\delta = 1$) and has no leaf vertices ($t = 0$), so Eq.~\eqref{eq:generic-count} predicts only the six homogeneous operations associated with the $H$ group. However, the corresponding graph state possesses additional Clifford symmetries that are not captured by the generic $H$ and $B$ construction. In particular, $Q_3$ admits a transversal $S$ operation. This is consistent with the fact that its stabilizer structure can be represented as a doubly-even CSS code~\cite{BravyiCross2015}, i.e., a CSS code in which every $X$-type stabilizer has Hamming weight divisible by $4$. This additional algebraic constraint enables transversal $S$ gate that preserve the stabilizer group but are not generated by the generic $H$ and $B$ mechanisms.

The existence of such additional symmetries does not limit the applicability of our framework. 
All simulation and optimization results below use the generic $H/B$ gate set~\eqref{eq:generic-count}, which is well-defined for arbitrary graphs and provides the operational search space for the noisy finite-size distillation circuits studied here.

\section{Graph-State Distillation and Circuit Optimization}
\label{sec4}
In the preceding sections, we developed a framework for graph-preserving Clifford operations, including their decomposition into homogeneous ($H$) and bilocal ($B$) components, their unification across LC orbits via minimum edge representatives (MERs), and their representation as permutations over a finite graph-state basis. This formulation leads to an exact and highly efficient simulation method, in which the action of a circuit can be evaluated by tracking permutations rather than evolving full stabilizer tableaux.

This reduction has direct algorithmic consequences. The restriction to graph-preserving operations yields a small, structured gate set tailored to the underlying graph, and each circuit instruction is evaluated through cached graph-basis label updates. The expensive step of deriving the action of admissible Clifford factors on the graph basis is performed once for the chosen graph representative, after which circuit optimization proceeds using compact symbolic gates rather than expanded stabilizer tableaux. Combined with the permutation-based simulation, this makes it possible to evaluate large numbers of candidate purification circuits exactly and efficiently under realistic noise models.

In this section, we leverage these properties to formulate graph-state distillation as a circuit-level optimization problem. Instead of designing asymptotic purification protocols, we explicitly search over finite-depth circuits acting on a small number of copies, incorporating hardware-specific noise and architectural constraints. This enables the systematic discovery of distillation circuits that are optimized for the given graph state and physical error model, and that can outperform asymptotically optimal constructions (e.g.\ the hashing method~\cite{Dur2003TwoColorable,Kruszynska2006AllGraph,Goyal2006}) in the regime of realistic finite-size noisy hardware.

\subsection{Overview of Graph-State Distillation}
Entanglement purification for multipartite graph states has been developed in several complementary directions over the past two decades~\cite{Dur2003TwoColorable,Kruszynska2006AllGraph}. These approaches form the conceptual background against which we position the graph-preserving circuit framework introduced in this work. We briefly summarize the main families of protocols and their scope.

\paragraph*{Two-colorable graph states as the foundational case.}
For all bipartite (two-colorable) graph states, Aschauer, Dür, and Briegel~\cite{Dur2003TwoColorable} constructed explicit LOCC recurrence protocols as well as hashing/breeding schemes with nonzero asymptotic yield.
These methods exploit the CSS structure of two-colorable graphs: local CNOTs between color classes separate $X$-type and $Z$-type stabilizer information, enabling parity checks and syndrome extraction via local measurements.
Subsequent work~\cite{Goyal2006} refined the analysis, providing scalable recurrence relations and improved hashing protocols.
Two-colorable graph states therefore serve as the analytically tractable "base case" for multipartite purification theory.

\paragraph*{Extensions to arbitrary graph states.}
For general $k$-colorable graphs, Kruszynska et al.~\cite{Kruszynska2006AllGraph} showed that two-colorable auxiliary graph states can be generated from noisy copies of the target state via LOCC.
These auxiliary states act as syndrome extractors that enable the same parity-check mechanisms used in the two-colorable case.
The resulting recurrence and breeding protocols apply to \emph{all} graph states (up to local Clifford equivalence), with fidelity thresholds and yields comparable to those obtained in the bipartite case.  
This establishes that arbitrary graph states can be distilled without relying on any particular application or geometry.

\paragraph*{Purification of arbitrary stabilizer states.}
More generally, Glancy et al.~\cite{glancy2006} formulated multipartite purification as a form of quantum error correction applied across many copies of a target stabilizer state.
Each party measures stabilizers of a chosen quantum code on their local block of copies, exchanges classical syndromes, applies corrections, and decodes.
This perspective unifies entanglement purification with stabilizer QEC and gives a fully general route for purifying \emph{any} stabilizer state, including all graph states, provided a suitable code is chosen.

\paragraph*{Position of the present work.}
The protocols above are LOCC-based and operate by extracting stabilizer parities from many noisy copies, typically analyzed in asymptotic or large-block regimes.  
In contrast, the framework developed in this paper focuses on \emph{circuit-level} transformations acting on a small number of copies (primarily two at a time), with the explicit goal of constructing optimized distillation circuits under realistic noise models.

Rather than designing universal purification protocols with asymptotic guarantees, we formulate distillation as a finite-size circuit optimization problem. By identifying the graph-preserving Clifford operations---determined by bipartiteness and leaf structure, and unified across the LC orbit via the MER---we obtain a restricted but efficiently simulable gate set tailored to the given graph. This restriction makes it possible to explicitly search over admissible circuits while incorporating hardware-specific features such as gate-dependent noise, measurement errors, and connectivity constraints.

As a result, our approach is not intended to characterize asymptotic yields or thresholds in the limit of large numbers of copies. Instead, it targets noisy, register-constrained hardware and produces finite-size purification circuits that are optimized for a specified hardware noise model and outperform the standard asymptotic protocols.

\begin{figure}[htbp]
    \centering
    \includegraphics[width=\columnwidth]{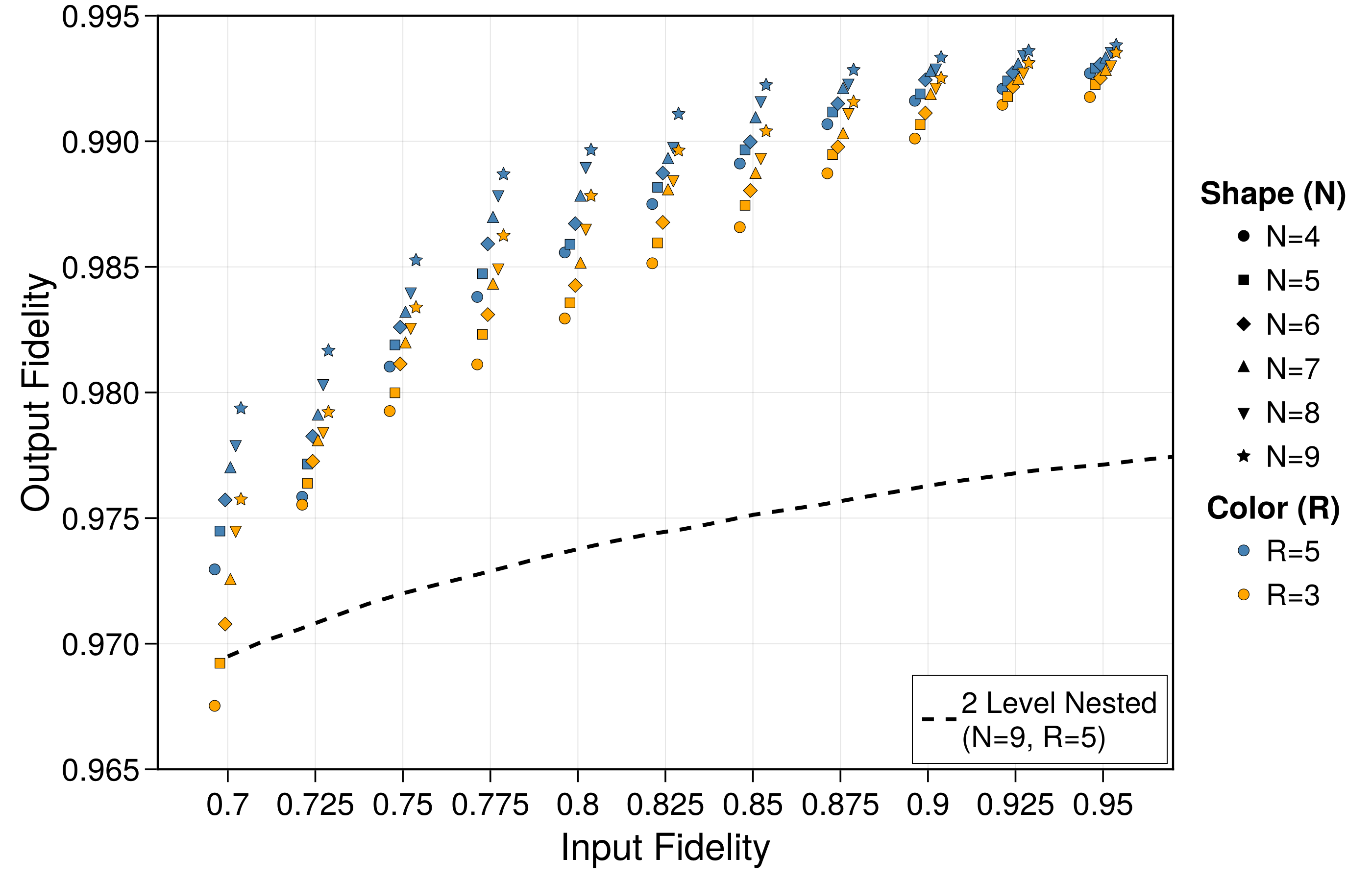}
    \caption{
    \textbf{Comparison between optimized graph-preserving circuits and the nested distillation protocol.}
    Output fidelity as a function of input fidelity for the five-qubit T-shaped graph state whose topology is shown in Fig.~\ref{fig:circuits}.
    Each point represents a circuit obtained by our optimization procedure under the specified parameters.
    Marker shapes indicate the number of raw graph states $N$ used by the circuit, while colors denote the register size $R$ (the maximum number of qubits that can be stored at a node simultaneously). 
    For visual clarity, a small horizontal offset is applied to the data points to avoid overlap; this does not affect the underlying values.
    The solid curve shows the performance of the two-level nested protocol using $N=9$ raw graph states.
    Executing the two-level nested protocol requires register size $R=5$.
    All circuits are optimized to maximize the output fidelity assuming gate error rate $p_2=0.01$ and measurement error rate $\eta=0.01$.
    For moderate input fidelities, the optimized circuits achieve higher output fidelity than the nested protocol even when using fewer raw states. 
    In addition, unlike nested protocols---which require fixed and exponentially growing numbers of raw states---our approach allows arbitrary input counts, enabling greater flexibility for circuit-level optimization.
    }
    \label{fig:2lv}
\end{figure}

\subsection{Noise Modeling for General Graph States}
In this work, the raw input states are $n$-qubit graph states subject to isotropic Pauli noise.
Given a graph $G$ with graph state $\ket{G}$, the noisy input state is represented as
\begin{equation}
\rho_\mathrm{in}
    \;=\;
    f_\mathrm{in}\,\ket{G}\!\bra{G}
    \;+\;
    (1-f_\mathrm{in})\,\frac{I_n-\ket{G}\!\bra{G}}{2^n-1},
\end{equation}
where $f_\mathrm{in}$ denotes the fidelity with respect to the ideal graph state, and $I_n$ is the
$n$-qubit identity operator.
This model captures the common situation in which graph states are generated through imperfect
entangling operations, local decoherence, or memory noise in distributed architectures.

\paragraph*{Depolarizing noise model.}
In our simulations, noise is associated with the failure of quantum operations and is modeled by depolarizing channels at the level of individual operations. We distinguish between single-qubit and two-qubit operations.

For a single-qubit operation, we denote the error probability by $p_1$. The corresponding noise channel is the standard single-qubit depolarizing channel,
\begin{equation}
\rho \;\longrightarrow\;
(1 - p_1)\rho
\;+\;
\frac{p_1}{3}\,(X\rho X + Y\rho Y + Z\rho Z),
\end{equation}
which applies each nontrivial Pauli error with equal probability.

For a two-qubit operation, we denote the error probability by $p_2$, interpreted as the failure probability of the gate as a whole. The corresponding channel is the two-qubit depolarizing channel,
\begin{equation}
\rho \;\longrightarrow\;
(1 - p_2)\rho
\;+\;
\frac{p_2}{15}\sum_{\substack{P \in \{I,X,Y,Z\}^{\otimes 2} \\ P \neq I\otimes I}}
P \rho P,
\end{equation}
which applies each nontrivial two-qubit Pauli error with equal probability.

All Clifford gates in this work--both homogeneous (\(H\)) and bilocal (\(B\)) operations--are modeled as ideal gates followed by the depolarizing channel corresponding to the gate arity. In the numerical simulations reported below, we set \(p_1=p_2\), so that the same error probability is used for both single- and two-qubit gates. Measurements are modeled as classical readout channels that flip the measurement outcome with probability \(\eta\). The parameters \(p_2\) and \(\eta\) therefore specify the gate and measurement noise used in the simulations.

Because graph states are stabilizer states, Pauli noise maps them to mixtures of graph-basis states. As a result, all noise processes considered here remain within the stabilizer formalism and can be simulated efficiently within the graph-preserving framework.

\paragraph*{Generality and biased noise.}
Although depolarizing noise is used for concreteness, the framework naturally extends to arbitrary
biased Pauli channels,
\begin{equation}
\mathcal{E}(\rho)
    =
    (1 - p_X - p_Y - p_Z)\rho
    + p_X X\rho X
    + p_Y Y\rho Y
    + p_Z Z\rho Z,
\end{equation}
which include physically important scenarios such as
dephasing-dominated noise (large $p_Z$) and amplitude-damping channels, the latter being well-approximated by Pauli twirling~\cite{zhou2013}.
Because graph states are stabilizer states, all such Pauli channels remain within the stabilizer formalism and are therefore fully supported by our simulation toolkit.

\paragraph*{Relevance for circuit optimization.}
The distillation and optimization procedures described in the next subsection aim to identify graph-preserving distillation circuits with high performance under the physical noise processes described above. All noise models considered here---state preparation noise, gate noise, and measurement noise---remain within the stabilizer formalism and can therefore be simulated efficiently. This allows systematic comparison of alternative graph-preserving strategies under realistic experimental conditions.

\begin{figure}[htbp]
    \centering
    \includegraphics[width=\columnwidth]{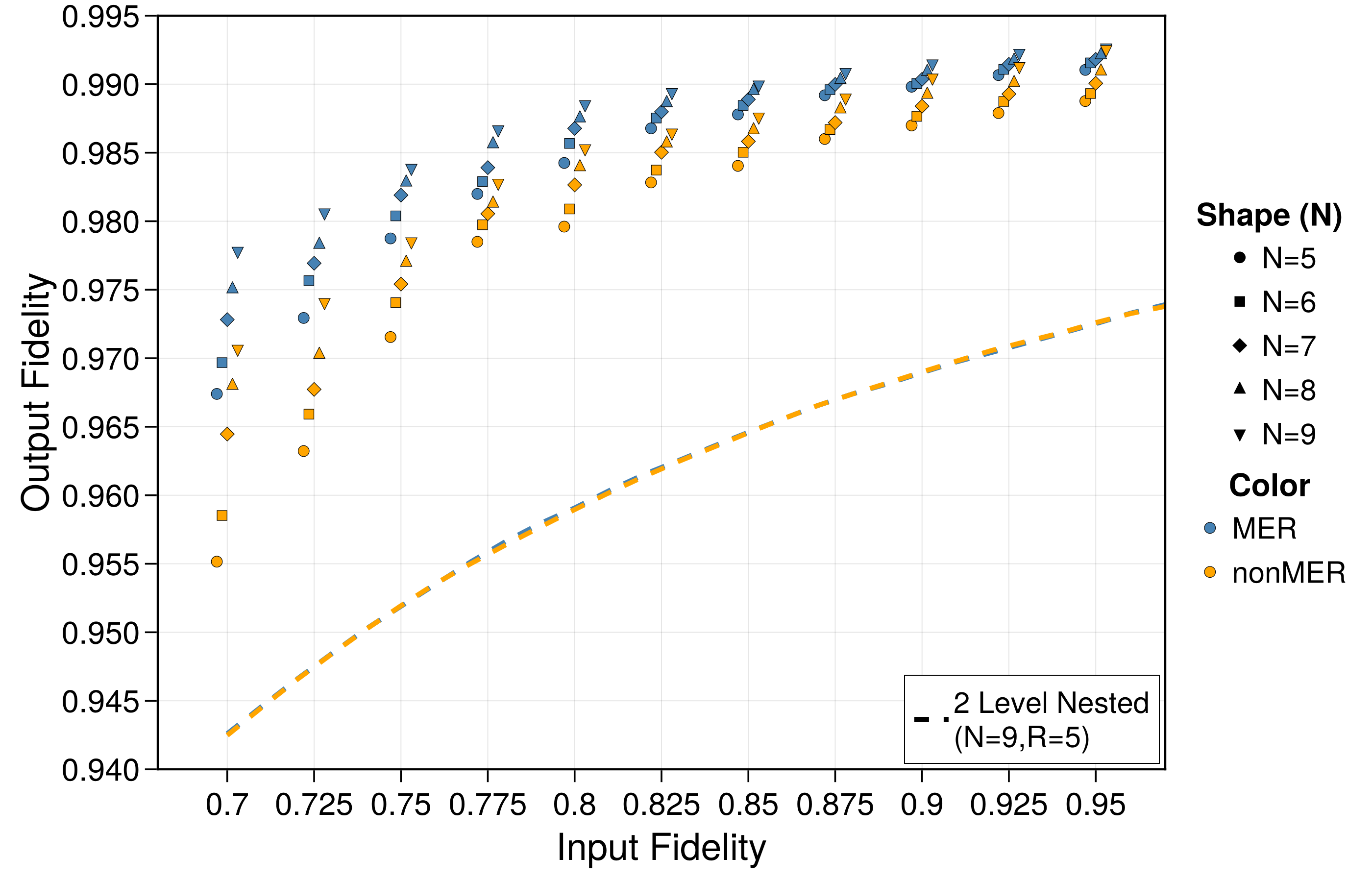}
    \caption{
    \textbf{Effect of graph representation on distillation performance within the same LC orbit.}
    Output fidelity as a function of input fidelity for optimized distillation circuits on the \(4\)-qubit linear graph (MER) and the LC-equivalent \(4\)-qubit cycle graph, together with the corresponding two-level nested-protocol baselines.
    Each point represents a circuit obtained by our optimization procedure under the specified parameters. Marker shapes indicate the number of raw graph states $N$ used by the circuit. The local register size is fixed to $R=5$, which is the minimum required to implement the two-level nested protocol used as a baseline. 
    For visual clarity, a small horizontal offset is applied to the data points to avoid overlap; this does not affect the underlying values.
    The two graph representations belong to the same LC orbit, as shown in Fig.~\ref{fig:lc-orbit-table}. Gate and measurement noise are set to $p_2=0.01$ and $\eta=0.01$, respectively.
    The MER admits a larger graph-preserving gate set than the non-MER, due to the additional bilocal ($B$-group) operations enabled by its leaf vertices. In the optimized circuits shown here, this enlarged gate set leads to higher output fidelity for the MER across the explored parameter range, with the separation most pronounced in the low-fidelity and small-$N$ regime, where the optimization is most constrained by the available gate set.
    The two nested-protocol baselines nearly coincide, indicating that the standard nested construction is essentially insensitive to the representative chosen within the LC orbit.
    Thus, representative choice affects the optimized finite-size circuits even when the underlying graph states are locally equivalent.
    The detailed distillation role of \(B\)-group operations is discussed in Appendix~\ref{app:bgroup_behavior}.
    }
    \label{fig:mer_vs_nonmer}
\end{figure}

\subsection{Circuit Optimization for Graph State Distillation}

Building on the graph-preserving gate framework developed in the previous sections, we perform numerical optimization of distillation circuits for general graph states. The central advantage of this framework is the drastic reduction of the circuit search space: by restricting attention to graph-preserving Clifford operations, the action of a circuit reduces to a permutation over a finite graph-state basis. This allows exact and efficient simulation of candidate circuits under realistic noise models and makes large-scale circuit optimization feasible.

For a given target graph state, the admissible gate set is determined by explicit graph-theoretic criteria---namely bipartiteness and the presence of leaf vertices---and may be further enlarged by considering an appropriate representative within the LC orbit. Regardless of whether one works directly with the original graph or with such a representative, the resulting gate set defines a well-structured and efficiently enumerable family of circuits suitable for automated search and optimization.

We carry out circuit optimization using a genetic algorithm, in which each candidate circuit is represented as a finite sequence of graph-preserving gates drawn from the admissible set. The specific choice of optimization heuristic is not essential; rather, it is the availability of fast, exact simulation that enables systematic exploration of the design space. To improve convergence, we incorporate a simulated-annealing mechanism in which a generation-dependent temperature controls the acceptance of suboptimal candidates. This allows the search to escape local optima in early stages and gradually concentrate on high-performance circuit designs.

In all numerical evaluations, a circuit configuration is specified by the following parameters:
\begin{itemize}
    \item \( G \): the adjacency matrix representing the input graph states;
    \item \( N \): the number of noisy input graph states consumed in each distillation attempt;
    \item \( K \): the number of output graph states produced upon successful distillation;
    \item \( R \): the size of the local quantum register available at each node, allowing for
    restricted local memory and regeneration of raw graph states during execution.
\end{itemize}

These parameters, together with the gate error rates \(p_2\), the measurement error rate \(\eta\), and the input fidelity \(f_{\mathrm{in}}\), define the operational constraints under which the optimization is performed. By varying these quantities, the same framework can model a wide range of experimental architectures, including memory-limited nodes and distributed quantum networks.

Each candidate circuit is evaluated according to user-defined performance metrics, such as output fidelity, success probability, or composite figures of merit balancing resource consumption against fidelity gain. The optimization procedure systematically explores the resulting design space and identifies circuits that outperform standard recurrence-based distillation strategies for the same target graph state (see Fig.~\ref{fig:2lv}).

The same optimization framework also enables a direct comparison between different graph representatives within a fixed LC orbit. In Fig.~\ref{fig:mer_vs_nonmer}, we compare optimized circuits constructed from a minimum-edge representative (MER) and a non-MER graph belonging to the same LC orbit. Since the MER has additional leaf vertices, it admits a larger graph-preserving gate set through the corresponding bilocal \(B\)-group operations. This enlarged admissible set gives the optimizer additional circuit-level freedom.

In the example shown, this advantage is modest overall but clearly visible in the low-fidelity, small-\(N\) regime, where the optimization is most constrained by the available gate set. As the number of raw states increases or the input fidelity becomes higher, the difference between MER and non-MER becomes small, indicating that the optimization is no longer strongly limited by the extra \(B\)-group freedom in that regime. By contrast, the two nested-protocol baselines for the MER and non-MER nearly coincide, showing that standard recurrence-style constructions are essentially insensitive to the choice of representative within the LC orbit.

These observations clarify the role of the MER in the present framework. Its importance is not that it always produces a dramatic fidelity improvement, but that it provides a canonical orbit-level representative that maximizes the accessible graph-preserving gate set. Any performance gain arising from representation-dependent gate availability can therefore only be accessed at the circuit-optimization level, not by fixed nested constructions.

Overall, this optimization approach provides a scalable and flexible route to graph-state distillation. By combining a principled restriction of admissible operations with efficient faster-than-stabilizer-based simulation, it enables the automated discovery of high-performance distillation circuits for arbitrary graph states under realistic noise conditions.

\section{Conclusion and Discussion}
\label{sec5}
In this work, we developed a unified framework for enumerating, simulating, and optimizing graph-preserving Clifford circuits for multipartite graph states.
By characterizing graph-preserving operations in terms of two elementary gate families---the homogeneous ($H$) group and the bilocal ($B$) group---we obtained a compact and efficiently enumerable gate set whose applicability is determined by simple graph-theoretic features.
Crucially, by organizing these operations across local-complementation (LC) orbits and anchoring the construction on minimum-edge representatives (MERs), we established a consistent assignment of admissible graph-preserving gates for all locally equivalent graph states.

This structural characterization enables an $O(1)$-cost stabilizer-level simulation of graph-preserving circuits.
Combined with this fast simulation capability, we introduced a circuit-level optimization framework that systematically searches for high-performance graph-state distillation circuits under realistic noise models.
Rather than relying on LOCC-based parity extraction or code-based purification, our approach operates directly at the circuit level, exploiting the intrinsic constraints imposed by the graph-state stabilizer structure.
The resulting circuits exhibit improved performance in terms of output fidelity, success probability, and resource efficiency, demonstrating that restricting to graph-preserving operations does not limit practical distillation power.

From a broader perspective, this work positions graph-preserving circuit design as a natural complement to existing multipartite entanglement purification frameworks.
While prior approaches emphasize generality through coding-theoretic or measurement-based techniques, the present framework isolates a graph-adapted subset of Clifford operations that can be simulated, enumerated, and optimized with minimal overhead.
This trade-off---reduced gate universality in exchange for strong structural compatibility and computational tractability---appears particularly well suited for near-term distributed quantum architectures, where locality, memory constraints, and noise robustness play a central role.

Several directions for future work remain open.
On the theoretical side, a complete combinatorial proof of the MER-based bipartiteness criterion would strengthen the classification of homogeneous graph-preserving operations across LC orbits, and may reveal deeper connections between local complementation, edge minimization, and graph
invariants.
Beyond the generic gate set identified here, exceptional graphs with additional Clifford symmetries merit a systematic classification, potentially linking graph-preserving operations to automorphism groups and CSS-compatible structures.
On the practical side, extending the present optimization framework to dynamical network models, time-dependent noise, or heterogeneous hardware constraints would further bridge the gap between abstract distillation protocols and real quantum-network deployments.

Overall, the framework introduced in this work provides a principled and scalable approach to graph-state distillation and circuit optimization.
By grounding circuit design in graph-theoretic structure and stabilizer-preserving symmetry, it opens a new pathway for constructing efficient, noise-robust entanglement-processing protocols for general graph states.
We expect these techniques to be broadly applicable to quantum networking, distributed quantum computation, and other settings where multipartite entanglement serves as a fundamental resource.

An open-source implementation of the graph-preserving simulation and optimization framework developed in this work is available at~\cite{github}.

\begin{acknowledgments}
We acknowledge support from NSF grants 1941583, 2346089, 2402861, 2522101. KG acknowledges the support from the Alexander von Humboldt Foundation. 
    
\end{acknowledgments}

\bibliography{bibliography}

\appendix

\section{Symplectic Representation of Clifford Operations}

Throughout this work we classify graph-preserving gates at the level of their action on stabilizer generators.
For this purpose we briefly recall the symplectic representation of the Clifford group.

The $n$-qubit Pauli group $\mathcal{P}_n$ is generated by single-qubit $X_i$ and $Z_i$ operators.
The Clifford group $\mathcal{C}_n$ is defined as the normalizer of $\mathcal{P}_n$ in $U(2^n)$, that is, unitaries that map Pauli operators to Pauli operators under conjugation.

Since Clifford operators are uniquely determined (up to global phase) by their action on a generating set of Pauli operators, it is natural to represent them via their induced linear action on the $2n$-dimensional binary vector space associated with Pauli strings.
Modding out by Pauli phases, one obtains the phaseless Clifford group
\begin{equation}
    \mathcal{C}_n^* = \mathcal{C}_n / \mathcal{P}_n ,
\end{equation}

which admits the symplectic representation
\begin{equation}
    \mathcal{C}_n^* \cong Sp(2n,\mathbb{F}_2).
\end{equation}

This representation allows Clifford gates to be treated as binary symplectic matrices acting on stabilizer generators, which forms the computational basis for the enumeration and group-structure analysis of the $H$ and $B$ subgroups introduced in the main text~\cite{koenig2014,ozols2008}.

Graph-state basis elements differ by local Pauli sign patterns on stabilizer generators. Since multiplying a Clifford by a local Pauli only changes the phase column of its tableau, we classify graph-preserving operations in the quotient $\mathcal{C}_2^*=\mathcal{C}_2/\mathcal{P}_2$. This preserves the induced permutation action on the graph-state basis while ignoring irrelevant local phases.

\section{Graph-State Stabilizers in Matrix Form}
\label{app:graph-stabilizer}

In this appendix we briefly review the stabilizer representation of graph states in a form suitable for the structural analysis of graph-preserving Clifford operations used in the main text.

\subsection{Graph-State Stabilizers}

Let $G=(V,E)$ be a simple graph with $|V|=n$ and adjacency matrix $\Gamma \in \mathbb{F}_2^{n\times n}$.
The corresponding graph state $|G\rangle$ is defined as the unique joint $+1$ eigenstate of the stabilizer generators
\begin{equation}
K_i \;=\; X_i \prod_{j \in N(i)} Z_j ,
\qquad i=1,\dots,n,
\end{equation}
where $N(i)$ denotes the neighborhood of vertex $i$.

In binary symplectic form, each stabilizer generator $K_i$ is represented by a pair of binary vectors
\begin{equation}
    (\mathbf{x}_i \mid \mathbf{z}_i) \in \mathbb{F}_2^{2n},
\end{equation}

where
\begin{equation}
    \mathbf{x}_i = e_i, 
    \qquad
    \mathbf{z}_i = \Gamma_{i,*}.
\end{equation}

Here $e_i$ is the $i$th standard basis vector and $\Gamma_{i,*}$ is the $i$th row of the adjacency matrix.

Thus the full stabilizer tableau can be written compactly as
\begin{equation}
S_G \;=\; 
\begin{pmatrix}
I_n \mid \Gamma
\end{pmatrix},
\end{equation}
where arithmetic is over $\mathbb{F}_2$.

\subsection{Two-Copy Representation}

In the distillation framework considered in this work, we act on two copies of a graph state.
The stabilizer generators for the two-copy system can be written as
\begin{equation}
    S_G^{(2)} =
    \begin{pmatrix}
    I_n & 0 \mid \Gamma & 0 \\
    0 & I_n \mid 0 & \Gamma
    \end{pmatrix},
\end{equation}

where the first block corresponds to copy 1 and the second to copy 2.

For each vertex $i$, the local Pauli support is therefore described by a pair of binary labels
\begin{equation}
    (x_i^{(1)}, x_i^{(2)} \mid z_i^{(1)}, z_i^{(2)}),
\end{equation}

which may be viewed as an element of $\mathbb{F}_2^{4}$.
This local two-copy structure is the natural setting in which phaseless two-copy Clifford operations act.

\subsection{Bipartite Structure and the Homogeneous Subgroup}

Suppose $G$ is bipartite with partition $V=A\cup B$ and adjacency matrix
\begin{equation}
    \Gamma=
    \begin{pmatrix}
    0 & \Gamma_{AB} \\
    \Gamma_{AB}^T & 0
    \end{pmatrix}.
\end{equation}

In the two-copy setting, each vertex carries a pair of binary labels
\begin{equation}
    (x_i^{(1)},x_i^{(2)} \mid z_i^{(1)},z_i^{(2)}) \in \mathbb{F}_2^4,
\end{equation}

and any phaseless two-qubit Clifford acting locally on that vertex induces
an invertible linear transformation on the two-copy label space
\begin{align}
    (z_i^{(1)},z_i^{(2)})^T \;\mapsto\; M (z_i^{(1)},z_i^{(2)})^T,
        \\
    (x_i^{(1)},x_i^{(2)})^T \;\mapsto\; M (x_i^{(1)},x_i^{(2)})^T,
\end{align}

for some $M \in GL(2,\mathbb{F}_2)$.

Since $|GL(2,\mathbb{F}_2)|=(2^2-1)(2^2-2)=6$,
there are six such invertible linear transformations.
These correspond precisely to the six phaseless homogeneous Clifford gates
generated by $\{ \mathrm{CNOT}_{12}, \mathrm{CNOT}_{21}, \mathrm{SWAP} \}$,
forming a group isomorphic to $GL(2,\mathbb{F}_2)\cong S_3\cong D_3$.

Graph preservation requires that the stabilizer relations
\begin{equation}
    x_A = \Gamma_{AB} z_B,
    \qquad
    x_B = \Gamma_{AB}^T z_A
\end{equation}

remain invariant under this copy-mixing action.
This holds when vertices in partition $A$ apply $M$
while vertices in partition $B$ apply the contragredient transform
\begin{equation}
    N=(M^{-1})^T.
\end{equation}

With this bipartitioned assignment, the adjacency coupling is preserved,
and the resulting six operations form the homogeneous subgroup $H$
appearing in the main text.
\subsection{Leaf Vertices and the Bilocal Subgroup}

Let $i$ be a leaf vertex with unique neighbor $j$.
Its stabilizer generator is
\begin{equation}
    K_i = X_i Z_j,
\end{equation}

while $K_j$ contains $Z_i$ as one of its factors.

In binary symplectic form, the row corresponding to $K_i$ has support only on
the local $x_i$ component and the $z_j$ component.
Consequently, any graph-preserving two-copy Clifford must map this row
to another stabilizer row of identical support pattern.
In particular:

\begin{itemize}
\item
The $z$-label of vertex $i$ cannot spread to other neighbors,
since $i$ has only one adjacent vertex.

\item
The pair $(z_i^{(1)},z_i^{(2)})$ may only mix within itself,
and any induced transformation must preserve the fact that $K_i$
contains exactly one $Z$ factor outside vertex $i$.

\item
The only admissible phaseless two-copy Clifford operations
consistent with these constraints are generated by:
    \begin{enumerate}
    \item local phase-type operations acting independently
          on the two-copy qubits of vertex $i$,
    \item a controlled-$Z$-type coupling between the two copies
          restricted to that vertex.
    \end{enumerate}
\end{itemize}

These generate an eight-element subgroup
\begin{equation}
    B_i \cong \mathbb{Z}_2^3.
\end{equation}

If $i$ and $k$ are distinct leaf vertices with disjoint neighborhoods,
their corresponding stabilizer supports do not overlap.
Therefore $B_i$ and $B_k$ act on disjoint subsystems and commute,
yielding a direct product
\begin{equation}
    \prod_{i \in \mathrm{Leaf}(G)} B_i.
\end{equation}

This structural locality is the origin of the $8^t$ factor
in the gate-counting formula of the main text.

\section{Local Complementation, LC Orbits, and Minimum-Edge Representatives}
\label{app:LC-MER}

\subsection{Local Complementation in Matrix Form}

Let $G=(V,E)$ be a graph with adjacency matrix $\Gamma \in \mathbb{F}_2^{n\times n}$, where arithmetic is performed over $\mathbb{F}_2$ and diagonal entries are zero.

For a vertex $v\in V$, denote by $N(v)$ its neighborhood. The local complementation (LC) at vertex $v$ transforms $G$ into a new graph $G^v$ whose adjacency matrix $\Gamma^v$ is given by

\begin{equation}
    \Gamma^v_{ij} =
        \begin{cases}
        \Gamma_{ij} + 1 & \text{if } i,j \in N(v),\ i\neq j,\\
        \Gamma_{ij} & \text{otherwise}.
        \end{cases}
\end{equation}

Equivalently, in matrix form one may write

\begin{equation}
    \Gamma^v
    =
    \Gamma
    +
    \Gamma_{*,v}\Gamma_{v,*}
    +
    \mathrm{diag}\!\left(\Gamma_{*,v}\Gamma_{v,*}\right)
    \pmod 2
\end{equation}

where $\Gamma_{*,v}$ denotes the $v$th column of $\Gamma$, and the diagonal term cancels. It follows immediately that $(G^v)^v = G$, so each LC operation is an involution.

\subsection{LC and Local Clifford Equivalence}

For graph states, local complementation corresponds to conjugation by a local Clifford unitary of the form

\begin{equation}
U_v = \sqrt{X_v} \bigotimes_{u\in N(v)} \sqrt{Z_u}.
\end{equation}

This operation maps the graph-state stabilizer
\begin{equation}
    S_G = (I \mid \Gamma)
\end{equation}
to
\begin{equation}
    S_{G^v} = (I \mid \Gamma^v),
\end{equation}

up to row permutations.
Therefore two graphs lie in the same LC orbit if and only if the corresponding graph states are locally Clifford equivalent (up to global phase).

For completeness, we note that the existence of a minimum-edge representative is immediate: for fixed $n$, there are only $2^{\binom{n}{2}}$ simple graphs on $n$ vertices, and LC operations map simple graphs to simple graphs. Hence every LC orbit is finite, and the edge count attains a minimum on it.

\subsection{Isomorphic LC Orbits}

Two graphs are isomorphic if their adjacency matrices satisfy
\begin{equation}
    \Gamma' = P \Gamma P^T
\end{equation}

for some permutation matrix $P$. Isomorphic graphs correspond to relabelings of qubits and therefore represent physically equivalent graph states.

The \emph{isomorphic LC orbit} is defined as the quotient of the labelled LC orbit under graph isomorphism. 
All classifications in the main text are therefore performed at the level of isomorphic LC orbits, which removes redundancies arising from vertex relabeling.

\subsection{Minimum-Edge Representatives}

Given a graph $G$, define

\begin{equation}
\mathrm{MER}(G)
=
\arg\min_{G' \in \mathcal{O}_{\mathrm{LC}}(G)} |E(G')|.
\end{equation}

Since the LC orbit is finite, $\mathrm{MER}(G)$ is nonempty.

\begin{figure}[t]
    \centering
    \includegraphics[width=0.95\linewidth]{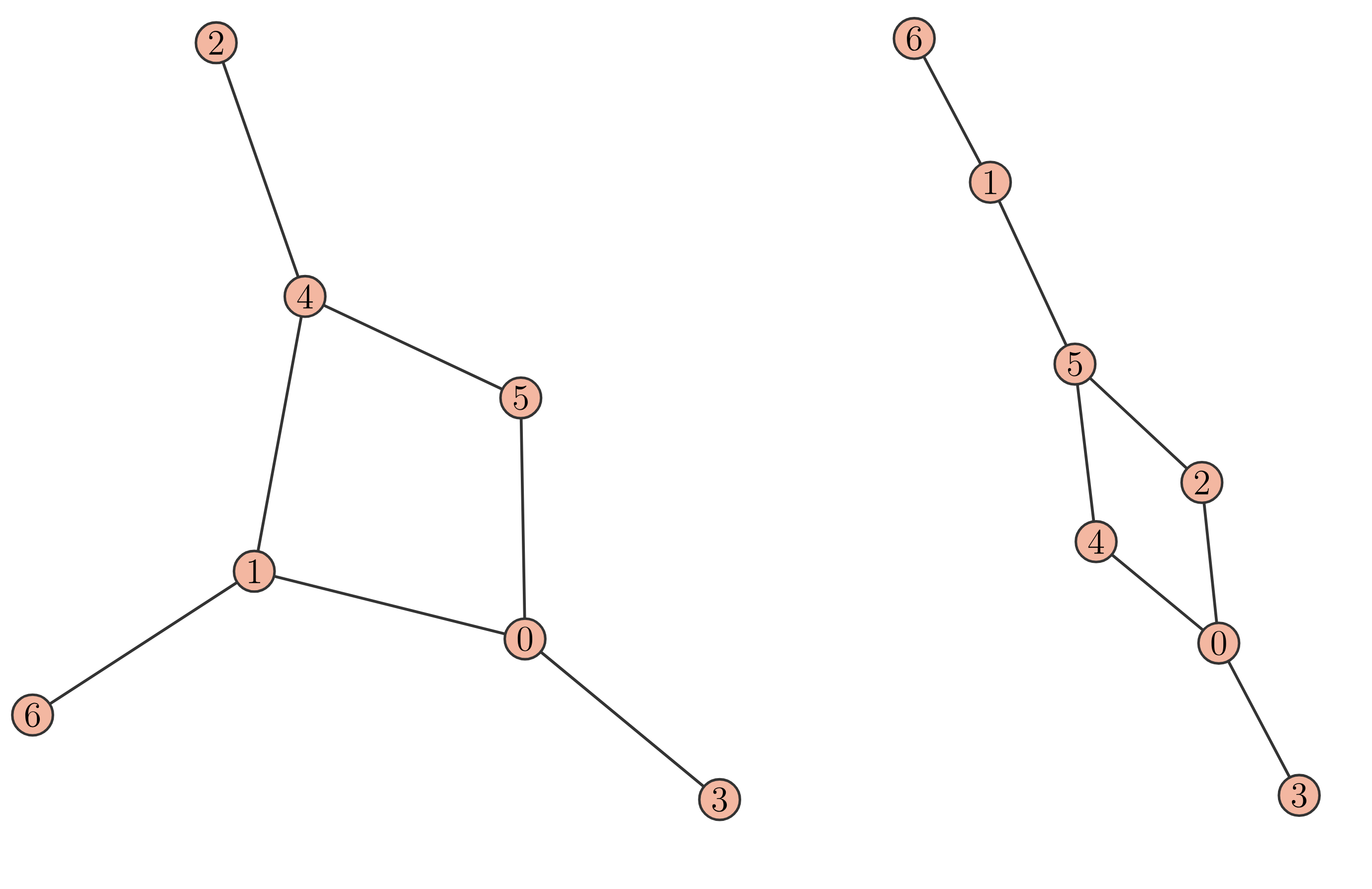}
    \caption{Two non-isomorphic $7$-vertex graphs lying in the same LC orbit, both with $|E|=7$ edges. They represent distinct labelled configurations within the same LC class and both qualify as minimum-edge representatives. Applying the LC sequence on vertices $(4,5,4)$ transforms one graph into the other.}
    \label{fig:non_unique}
\end{figure}

The minimum-edge representative within an LC orbit need not be unique.
Figure~\ref{fig:non_unique} shows an explicit example for $n=7$ vertices: two graphs that

\begin{itemize}
    \item lie in the same LC orbit,
    \item are non-isomorphic as graphs,
    \item and both attain the same minimum edge count $|E|=7$.
\end{itemize}

These graphs therefore both belong to $\mathrm{MER}(G)$.

This example illustrates that the MER construction selects a
\emph{set} of sparsest representatives rather than a unique graph.
When a unique representative is required, we select a canonical element from \( \mathrm{MER}(G) \) using the canonical indexing scheme introduced in Ref.~\cite{Cabello2011}.

\subsection{LC-Orbit Database}
Explicit LC-orbit enumerations have been carried out for small numbers of qubits. In this work we make use of a database of LC orbits constructed by exhaustive enumeration up to $n\le 12$ vertices.
For larger system sizes we refer to the LC-orbit databases available in the literature, which extend the classification to $n\le 16$ vertices~\cite{Sharma2026minimisingnumberof}.

\begin{table*}[t]
\begin{ruledtabular}
\begin{tabular}{ccccc|ccccc}

\textbf{EC} & \textbf{$n$} & \textbf{Orbit} & \textbf{MER} & \textbf{Max} &
\textbf{EC} & \textbf{$n$} & \textbf{Orbit} & \textbf{MER} & \textbf{Max} \\
            &              & \textbf{size}  & \textbf{leaves} & \textbf{leaves} &
            &              & \textbf{size}  & \textbf{leaves} & \textbf{leaves} \\
\hline
 3 & 4 &   2 & 3 & 3 &  26 & 7 &  16 & 4 & 4 \\
 4 & 4 &   4 & 2 & 2 &  27 & 7 &  44 & 3 & 3 \\
\hline
 5 & 5 &   2 & 4 & 4 &  28 & 7 &  44 & 3 & 3 \\
 6 & 5 &   6 & 3 & 3 &  29 & 7 &  14 & 3 & 3 \\
 7 & 5 &  10 & 2 & 2 &  30 & 7 &  66 & 2 & 2 \\
 8 & 5 &   3 & 0 & 0 &  31 & 7 &  10 & 3 & 3 \\
\hline
 9 & 6 &   2 & 5 & 5 &  32 & 7 &  10 & 4 & 4 \\
10 & 6 &   6 & 4 & 4 &  33 & 7 &  21 & 2 & 2 \\
11 & 6 &   4 & 4 & 4 &  34 & 7 &  26 & 3(2)* & 3 \\
12 & 6 &  16 & 3 & 3 &  35 & 7 &  36 & 2 & 2 \\
13 & 6 &  10 & 3 & 3 &  36 & 7 &  28 & 3 & 3 \\
14 & 6 &  25 & 2 & 2 &  37 & 7 &  72 & 2 & 2 \\
15 & 6 &   5 & 2 & 2 &  38 & 7 & 114 & 1 & 1 \\
16 & 6 &   5 & 3 & 3 &  39 & 7 &  56 & 1 & 1 \\
17 & 6 &  21 & 1 & 1 &  40 & 7 &  92 & 0 & 0 \\
18 & 6 &  16 & 0 & 0 &  41 & 7 &  57 & 1 & 1 \\
19 & 6 &   2 & 0 & 0 &  42 & 7 &  33 & 0 & 0 \\
\hline
20 & 7 &   2 & 6 & 6 &  43 & 7 &   9 & 0 & 0 \\
21 & 7 &   6 & 5 & 5 &  44 & 7 &  46 & 0 & 0 \\
22 & 7 &   6 & 5 & 5 &  45 & 7 &   9 & 1 & 1 \\
23 & 7 &  16 & 4 & 4 &     &   &     &   &   \\
24 & 7 &  10 & 4 & 4 &     &   &     &   &   \\
25 & 7 &  10 & 4 & 4 &     &   &     &   &   \\
\end{tabular}
\end{ruledtabular}
\caption{Leaf vertex counts for LC orbits with 4--7 qubits. For more detailed data, see~\url{https://github.com/QuantumSavory/LCOrbits.jl}.
\emph{Equivalence class} follows the standard ordering of~\cite{Cabello2011}.
\emph{MER leaf count} is the number of leaf vertices (degree-1 vertices) in
the minimum edge representative of the orbit; \emph{max leaf in orbit} is the
maximum over all graphs in the orbit. * EC 34 has two MERs as shown in Fig.~\ref{fig:non_unique}}
\label{tab:lc_orbit_leaves}
\end{table*}

\section{Distillation behavior of a bilocal B group operation}
\label{app:bgroup_behavior}

The role of CNOT gates in entanglement distillation is well understood from standard recurrence and nested protocols, where they transfer parity information between copies and allow unfavorable error patterns to be filtered out by postselection. By contrast, for the bilocal \(B\) group operations introduced in the main text, their operational effect in distillation is less immediate. In particular, it is not obvious from the abstract group description why the paired \(CZ\) and \(XCX\) gates acting on a leaf vertex and its neighbor should improve fidelity after measurement and postselection.

To make this mechanism explicit, we consider here the simplest nontrivial example: the \(4\)-qubit linear graph state. This is the smallest graph in which a \(B\) group operation acts on a leaf vertex and its unique neighbor while leaving at least one additional qubit outside the acted-on pair. It therefore provides the minimal setting in which the effect of the bilocal operation on distillation can be seen without collapsing to a trivial two- or three-qubit case.

We enumerate the graph-basis mappings induced by a single \(B\) group operation acting on one leaf vertex and its neighbor across two copies of the \(4\)-qubit linear graph state. Concretely, we apply the paired \(XCX\) gate on the leaf qubit and the corresponding \(CZ\) gate on its neighboring qubit, leaving the other leaf--neighbor pair untouched. The full basis-state mapping is given in Tables~\ref{tab:b_group_operations_1} and~\ref{tab:b_group_operations_2}, while Table~\ref{tab:bgroup_phase_mapping} lists only the contributions that survive the subsequent measurement-and-postselection step and contribute to the output fidelity.

Since the graph has four qubits, its graph-state basis contains \(2^4=16\) basis states. In Table~\ref{tab:bgroup_phase_mapping}, each state is represented by a string of \(+\) and \(-\) signs indicating the phase pattern of the four stabilizer generators, equivalently the graph-basis label. The first state in each input pair denotes the copy to be purified, while the second denotes the sacrificed copy. We assume the standard isotropic graph-basis noise model: the target all \(+\) graph-basis state has probability \(F\), while each of the remaining \(15\) basis states has equal probability
\begin{equation}
    q=\frac{1-F}{15}.
\end{equation}

Throughout this analysis we assume ideal operations: the \(B\) group gate is perfect and the measurements are noiseless. In addition, for graph-state distillation one measures only one side of the bipartition. For the \(4\)-qubit linear graph, we take this measured set to be qubits \(1\) and \(3\).

The purpose of Table~\ref{tab:bgroup_phase_mapping} is therefore to identify which input pairs are mapped, after the bilocal gate and the prescribed measurements, to accepted outcomes for which the purified copy remains in the all \(+\) graph-basis state. By explicitly tallying these retained contributions and normalizing by the total acceptance probability, one finds that the output fidelity is
\begin{equation}
    F_{\mathrm{out}}
    =
    \frac{F^2+3Fq}{F^2+3Fq+60q^2}.
\end{equation}
Since
\begin{equation}
    \frac{F^2+3Fq}{F^2+3Fq+60q^2}>F
    \qquad
    \text{for } F>\frac1{16},
    \label{eq:4q}
\end{equation}

this bilocal \(B\) group step strictly increases the fidelity throughout the physically relevant regime. This simple \(4\)-qubit example illustrates the mechanism that contributes to the performance separation between the \(4\)-qubit linear MER and the \(4\)-qubit cycle non-MER shown in Fig.~\ref{fig:mer_vs_nonmer}.

This example makes clear that the \(B\) group does not merely enlarge the admissible gate set in an abstract algebraic sense. Operationally, its paired \(XCX/CZ\) action redistributes stabilizer-phase information across the two copies: phase information initially associated with the leaf qubit is mapped into phase information carried by the corresponding neighboring site in the opposite copy. 
As seen in Table~\ref{tab:bgroup_phase_mapping}, this redistribution is localized to the acted-on leaf--neighbor pair: the phase entries corresponding to the two qubits outside the \(B\)-group support remain unchanged, whereas the phase pattern on the two qubits touched by the bilocal operation is modified. 
As a result, phase patterns that are initially localized on the leaf become encoded in a form that can be accessed by the subsequent measurements and postselection. The bilocal operation therefore enhances the weight of the target state relative to the uniform background of erroneous graph-basis states. In this sense, the \(B\) group plays a role analogous to the more familiar parity-transfer action of CNOT-based distillation steps, but adapted to the leaf structure of the graph state.

The effect is especially pronounced in the present \(4\)-qubit linear example because the bilocal \(B\)-group step acts on one leaf vertex and its unique neighbor, which already constitute a large fraction of the entire graph. As a result, the corresponding leaf-specific distillation mechanism has a comparatively strong impact on the overall fidelity. For larger graph states, the portion of the system directly accessible to such \(B\)-group steps is smaller, and the resulting improvement is therefore generally less pronounced.

A similar but less extreme behavior can be seen for the \(5\)-qubit \(T\)-shaped graph state shown in Fig.~\ref{fig:circuits}. Consider a single \(B\)-group step acting only on leaf vertex \(1\) and its neighboring vertex \(2\), followed by measurements on qubits \(2\) and \(5\), which belong to the same bipartition subset. Under the same idealized assumptions of perfect gates and noiseless measurements, and using the isotropic graph-basis noise model with
\begin{equation}
    q=\frac{1-F}{31},
\end{equation}

one obtains
\begin{equation}
    F_{\mathrm{out}}
=
\frac{F^2+7Fq}{F^2+22Fq+233q^2}>F
\qquad
    \text{for } F>\frac1{32},
    \label{eq:5t}
\end{equation}

\begin{figure}[ht]
    \centering
    \includegraphics[width=\linewidth]{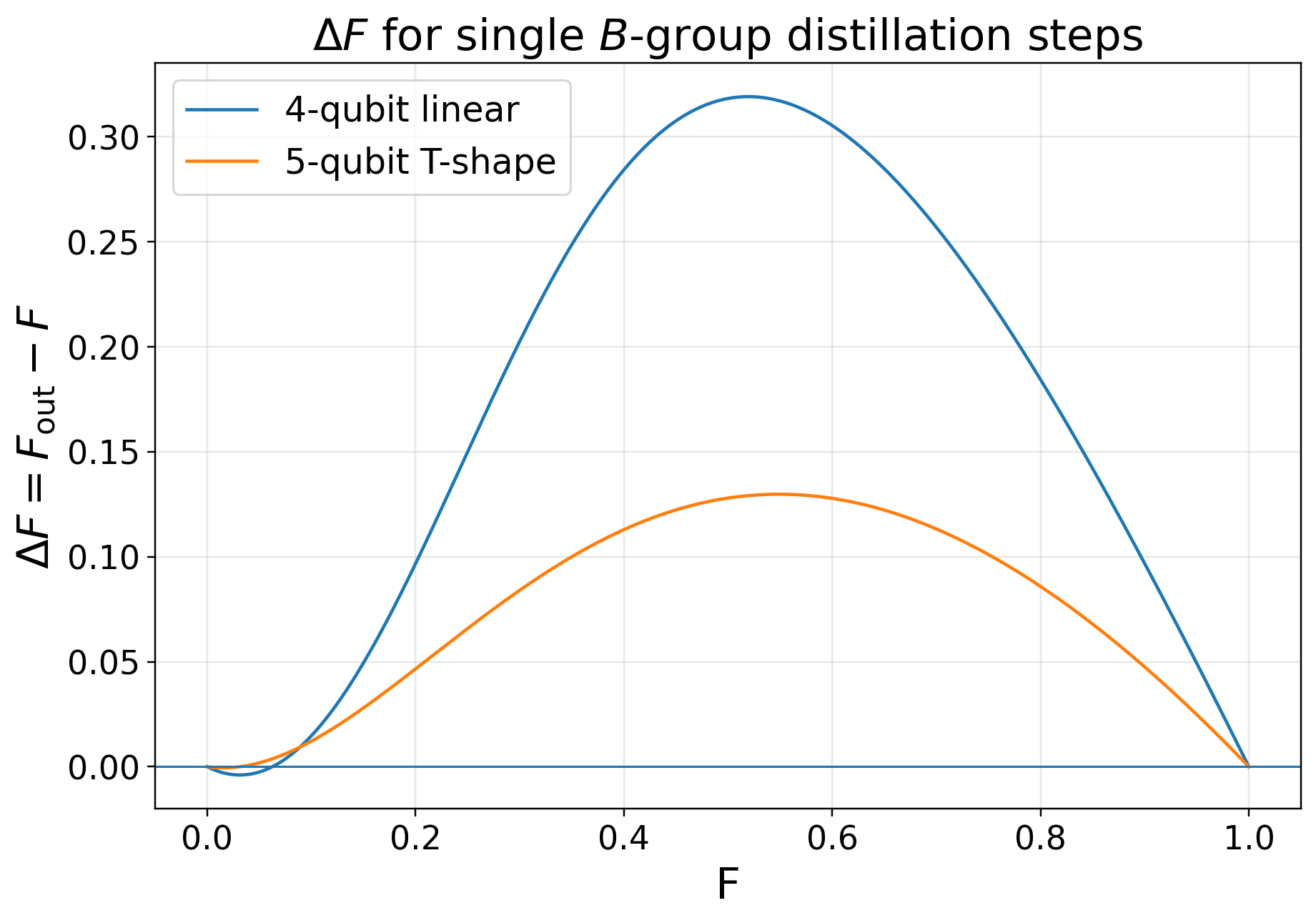}
    \caption{
    \textbf{Fidelity gain from a single \(B\)-group distillation step for two leaf-containing graph states.}
    The plot shows \(\Delta F = F_{\mathrm{out}}-F\) as a function of the input fidelity \(F\) for two examples discussed in Appendix~\ref{app:bgroup_behavior}: the \(4\)-qubit linear graph and the \(5\)-qubit \(T\)-shaped graph. In both cases, a single bilocal \(B\)-group operation strictly increases the fidelity throughout the physically relevant regime.
    The \(4\)-qubit linear graph exhibits a larger gain because the acted-on leaf--neighbor pair constitutes a larger fraction of the total graph, whereas for the \(5\)-qubit \(T\)-shaped graph the same \(B\)-group mechanism acts more locally and therefore yields a more moderate improvement.
    }
    \label{fig:bgroup_deltaF}
\end{figure}

so the \(B\)-group step again strictly increases the fidelity throughout the physically relevant regime. The improvement is, however, less pronounced than for the \(4\)-qubit linear graph. This is consistent with the interpretation above: in the \(5\)-qubit \(T\)-shaped graph, a single \(B\)-group step acts only on one leaf--neighbor pair and therefore performs only a more partial form of distillation on the full graph. 

The comparison between the two examples is shown explicitly in Fig.~\ref{fig:bgroup_deltaF}, where we plot the fidelity gain \(\Delta F = F_{\mathrm{out}}-F\) for the \(4\)-qubit linear and \(5\)-qubit \(T\)-shaped graph states. These examples illustrate a more general trend: even for larger graph states, a single \(B\)-group step still yields a strictly positive fidelity gain, but that gain becomes progressively smaller as the total number of qubits increases, since the underlying leaf--neighbor distillation mechanism remains local while the graph itself becomes larger.

The importance of \(B\)-group operations may be amplified under biased or spatially nonuniform noise. Because a \(B\)-group operation acts only on a leaf--neighbor pair, it can selectively transform error components concentrated on that pair without requiring a homogeneous operation across the full graph. In modular and networked architectures, leaf qubits may be deliberately assigned as communication or interface qubits that connect a local resource state to another module. Such qubits can undergo repeated entanglement-generation, initialization, reset, and measurement operations, leading to dephasing and crosstalk mechanisms distinct from those affecting storage qubits in the interior of the module~\cite{Reiserer2016Robust,Feng2024Crosstalk}. When this architectural role is mapped onto a leaf--neighbor pair, the resulting nonuniform error distribution may align particularly well with the support of the \(B\)-group action. The \(B\) group may therefore provide a larger advantage under such biased noise models than in the isotropic setting considered here.
The same interface assignment may also provide an operational advantage: a \(B\)-group operation is confined to the designated interface qubit and its neighbor, rather than requiring coordinated control across the full graph. When leaf vertices are mapped to such accessible interfaces, \(B\)-group distillation may therefore be favorable both in its noise selectivity and in its implementation requirements.

\begin{table*}[t]
\centering
\scriptsize
\setlength{\tabcolsep}{2.2pt}

\begin{minipage}[t]{0.49\textwidth}
\centering
\begin{ruledtabular}
\begin{tabular}{cccc}
Initial state & After B & Initial prob. & After prob. \\
\hline
$(++++,++++)$ & $(++++,++++)$ & $F^2$ &$F^2$ \\
$(++++,+++-)$ & $(++++,+++-)$ & $Fq$ &$Fq$ \\
$(++++,+-++)$ & $(++++,+-++)$ & $Fq$ & $Fq$\\
$(++++,+-+-)$ & $(++++,+-+-)$ & $Fq$ & $Fq$ \\
$(+++-,++++)$ & $(+++-,++++)$ & $qF$&$qF$ \\
$(+++-,+++-)$ & $(+++-,+++-)$ & $q^2$&$q^2$ \\
$(+++-,+-++)$ & $(+++-,+-++)$ & $q^2$&$q^2$ \\
$(+++-,+-+-)$ & $(+++-,+-+-)$ & $q^2$& $q^2$\\
$(++-+,++++)$ & $(++-+,++++)$ & $qF$&$qF$ \\
$(++-+,+++-)$ & $(++-+,+++-)$ & $q^2$&$q^2$ \\
$(++-+,+-++)$ & $(++-+,+-++)$ & $q^2$&$q^2$ \\
$(++-+,+-+-)$ & $(++-+,+-+-)$ & $q^2$&$q^2$ \\
$(++--,++++)$ & $(++--,++++)$ & $qF$& $qF$\\
$(++--,+++-)$ & $(++--,+++-)$ & $q^2$&$q^2$ \\
$(++--,+-++)$ & $(++--,+-++)$ & $q^2$&$q^2$ \\
$(++--,+-+-)$ & $(++--,+-+-)$ & $q^2$&$q^2$ \\
$(+-++,++++)$ & $(+-++,++++)$ & $qF$&$qF$ \\
$(+-++,+++-)$ & $(+-++,+++-)$ & $q^2$&$q^2$ \\
$(+-++,+-++)$ & $(+-++,+-++)$ & $q^2$&$q^2$ \\
$(+-++,+-+-)$ & $(+-++,+-+-)$ & $q^2$&$q^2$ \\
$(+-+-,++++)$ & $(+-+-,++++)$ & $qF$&$qF$ \\
$(+-+-,+++-)$ & $(+-+-,+++-)$ & $q^2$&$q^2$ \\
$(+-+-,+-++)$ & $(+-+-,+-++)$ & $q^2$&$q^2$ \\
$(+-+-,+-+-)$ & $(+-+-,+-+-)$ & $q^2$&$q^2$ \\
$(+--+,++++)$ & $(+--+,++++)$ & $qF$&$qF$ \\
$(+--+,+++-)$ & $(+--+,+++-)$ & $q^2$&$q^2$ \\
$(+--+,+-++)$ & $(+--+,+-++)$ & $q^2$&$q^2$ \\
$(+--+,+-+-)$ & $(+--+,+-+-)$ & $q^2$&$q^2$ \\
$(+---,++++)$ & $(+---,++++)$ & $qF$&$qF$ \\
$(+---,+++-)$ & $(+---,+++-)$ & $q^2$&$q^2$ \\
$(+---,+-++)$ & $(+---,+-++)$ & $q^2$&$q^2$ \\
$(+---,+-+-)$ & $(+---,+-+-)$ & $q^2$&$q^2$ \\
\end{tabular}
\end{ruledtabular}
\end{minipage}
\hfill
\begin{minipage}[t]{0.49\textwidth}
\centering
\begin{ruledtabular}
\begin{tabular}{cccc}
Initial state & After B & Initial prob. & After prob. \\
\hline
$(-+++,++++)$ & $(-+++,+-++)$ & $qF$&$q^2$ \\
$(-+++,+++-)$ & $(-+++,+-+-)$ & $q^2$&$q^2$ \\
$(-+++,+-++)$ & $(-+++,++++)$ & $q^2$&$qF$ \\
$(-+++,+-+-)$ & $(-+++,+++-)$ & $q^2$&$q^2$ \\
$(-++-,++++)$ & $(-++-,+-++)$ & $qF$&$q^2$ \\
$(-++-,+++-)$ & $(-++-,+-+-)$ & $q^2$&$q^2$ \\
$(-++-,+-++)$ & $(-++-,++++)$ & $q^2$&$qF$ \\
$(-++-,+-+-)$ & $(-++-,+++-)$ & $q^2$&$q^2$ \\
$(-+-+,++++)$ & $(-+-+,+-++)$ & $qF$&$q^2$ \\
$(-+-+,+++-)$ & $(-+-+,+-+-)$ & $q^2$&$q^2$ \\
$(-+-+,+-++)$ & $(-+-+,++++)$ & $q^2$&$qF$ \\
$(-+-+,+-+-)$ & $(-+-+,+++-)$ & $q^2$&$q^2$ \\
$(-+--,++++)$ & $(-+--,+-++)$ & $qF$&$q^2$ \\
$(-+--,+++-)$ & $(-+--,+-+-)$ & $q^2$&$q^2$ \\
$(-+--,+-++)$ & $(-+--,++++)$ & $q^2$&$qF$ \\
$(-+--,+-+-)$ & $(-+--,+++-)$ & $q^2$&$q^2$ \\
$(--++,++++)$ & $(--++,+-++)$ & $qF$&$q^2$ \\
$(--++,+++-)$ & $(--++,+-+-)$ & $q^2$&$q^2$ \\
$(--++,+-++)$ & $(--++,++++)$ & $q^2$&$qF$ \\
$(--++,+-+-)$ & $(--++,+++-)$ & $q^2$&$q^2$ \\
$(--+-,++++)$ & $(--+-,+-++)$ & $qF$&$q^2$ \\
$(--+-,+++-)$ & $(--+-,+-+-)$ & $q^2$&$q^2$ \\
$(--+-,+-++)$ & $(--+-,++++)$ & $q^2$&$qF$ \\
$(--+-,+-+-)$ & $(--+-,+++-)$ & $q^2$&$q^2$ \\
$(---+,++++)$ & $(---+,+-++)$ & $qF$&$q^2$ \\
$(---+,+++-)$ & $(---+,+-+-)$ & $q^2$&$q^2$ \\
$(---+,+-++)$ & $(---+,++++)$ & $q^2$&$qF$ \\
$(---+,+-+-)$ & $(---+,+++-)$ & $q^2$&$q^2$ \\
$(----,++++)$ & $(----,+-++)$ & $qF$&$q^2$ \\
$(----,+++-)$ & $(----,+-+-)$ & $q^2$&$q^2$ \\
$(----,+-++)$ & $(----,++++)$ & $q^2$&$qF$ \\
$(----,+-+-)$ & $(----,+++-)$ & $q^2$&$q^2$ \\
\end{tabular}
\end{ruledtabular}
\end{minipage}
\caption{
\textbf{Postselected contributions to the output fidelity for a single \(B\)-group step on the \(4\)-qubit linear graph state.}
The table contains only those graph-basis input pairs that are accepted by the measurement-and-postselection rule after applying one paired \(B\)-group operation to a leaf vertex and its unique neighbor across two copies of the \(4\)-qubit linear graph state. It does not show the full basis-state mapping; the complete mapping is given in Tables~\ref{tab:b_group_operations_1} and~\ref{tab:b_group_operations_2}. The \(+\)/\(-\) strings denote stabilizer-phase patterns, the first state in each pair is the purified copy, and the second is the sacrificed copy. Under the isotropic graph-basis noise model, the all \(+\) state has probability \(F\) and each other basis state has probability \(q=(1-F)/15\). The listed contributions assume ideal gates and noiseless measurements on qubits \(1\) and \(3\) of the sacrificed copy, which form one side of the bipartition of the \(4\)-qubit linear graph; as in standard graph-state distillation protocols, postselection is based on local \(X\)- or \(Z\)-basis measurements performed only on one bipartition subset.
}
\label{tab:bgroup_phase_mapping}
\end{table*}

\begin{table*}[p]
\caption{\label{tab:b_group_operations_1} Full mapping of a \(B\)-group operation on one leaf of the \(4\)-qubit linear state (Part I).}
\renewcommand{\arraystretch}{0.90}
\begin{ruledtabular}
\begin{tabular}{llll}
\multicolumn{2}{c}{Part I-a} & \multicolumn{2}{c}{Part I-b} \\
\hline
Initial state & After B group operation &
Initial state & After B group operation \\
\hline
$(++++,++++)$ & $(++++,++++)$ & $(+-++,++++)$ & $(+-++,++++)$ \\
$(++++,+++-)$ & $(++++,+++-)$ & $(+-++,+++-)$ & $(+-++,+++-)$ \\
$(++++,++-+)$ & $(++++,++-+)$ & $(+-++,++-+)$ & $(+-++,++-+)$ \\
$(++++,++--) $ & $(++++,++--) $ & $(+-++,++--) $ & $(+-++,++--) $ \\
$(++++,+-++)$ & $(++++,+-++)$ & $(+-++,+-++)$ & $(+-++,+-++)$ \\
$(++++,+-+-)$ & $(++++,+-+-)$ & $(+-++,+-+-)$ & $(+-++,+-+-)$ \\
$(++++,+--+)$ & $(++++,+--+)$ & $(+-++,+--+)$ & $(+-++,+--+)$ \\
$(++++,+---)$ & $(++++,+---)$ & $(+-++,+---)$ & $(+-++,+---)$ \\
$(++++,-+++)$ & $(+-++,-+++)$ & $(+-++,-+++)$ & $(++++,-+++)$ \\
$(++++,-++-)$ & $(+-++,-++-)$ & $(+-++,-++-)$ & $(++++,-++-)$ \\
$(++++,-+-+)$ & $(+-++,-+-+)$ & $(+-++,-+-+)$ & $(++++,-+-+)$ \\
$(++++,-+--)$ & $(+-++,-+--)$ & $(+-++,-+--)$ & $(++++,-+--)$ \\
$(++++,--++)$ & $(+-++,--++)$ & $(+-++,--++)$ & $(++++,--++)$ \\
$(++++,--+-)$ & $(+-++,--+-)$ & $(+-++,--+-)$ & $(++++,--+-)$ \\
$(++++,---+)$ & $(+-++,---+)$ & $(+-++,---+)$ & $(++++,---+)$ \\
$(++++,----)$ & $(+-++,----)$ & $(+-++,----)$ & $(++++,----)$ \\
$(+++-,++++)$ & $(+++-,++++)$ & $(+-+-,++++)$ & $(+-+-,++++)$ \\
$(+++-,+++-)$ & $(+++-,+++-)$ & $(+-+-,+++-)$ & $(+-+-,+++-)$ \\
$(+++-,++-+)$ & $(+++-,++-+)$ & $(+-+-,++-+)$ & $(+-+-,++-+)$ \\
$(+++-,++--) $ & $(+++-,++--) $ & $(+-+-,++--) $ & $(+-+-,++--) $ \\
$(+++-,+-++)$ & $(+++-,+-++)$ & $(+-+-,+-++)$ & $(+-+-,+-++)$ \\
$(+++-,+-+-)$ & $(+++-,+-+-)$ & $(+-+-,+-+-)$ & $(+-+-,+-+-)$ \\
$(+++-,+--+)$ & $(+++-,+--+)$ & $(+-+-,+--+)$ & $(+-+-,+--+)$ \\
$(+++-,+---)$ & $(+++-,+---)$ & $(+-+-,+---)$ & $(+-+-,+---)$ \\
$(+++-,-+++)$ & $(+-+-,-+++)$ & $(+-+-,-+++)$ & $(+++-,-+++)$ \\
$(+++-,-++-)$ & $(+-+-,-++-)$ & $(+-+-,-++-)$ & $(+++-,-++-)$ \\
$(+++-,-+-+)$ & $(+-+-,-+-+)$ & $(+-+-,-+-+)$ & $(+++-,-+-+)$ \\
$(+++-,-+--)$ & $(+-+-,-+--)$ & $(+-+-,-+--)$ & $(+++-,-+--)$ \\
$(+++-,--++)$ & $(+-+-,--++)$ & $(+-+-,--++)$ & $(+++-,--++)$ \\
$(+++-,--+-)$ & $(+-+-,--+-)$ & $(+-+-,--+-)$ & $(+++-,--+-)$ \\
$(+++-,---+)$ & $(+-+-,---+)$ & $(+-+-,---+)$ & $(+++-,---+)$ \\
$(+++-,----)$ & $(+-+-,----)$ & $(+-+-,----)$ & $(+++-,----)$ \\
$(++-+,++++)$ & $(++-+,++++)$ & $(+--+,++++)$ & $(+--+,++++)$ \\
$(++-+,+++-)$ & $(++-+,+++-)$ & $(+--+,+++-)$ & $(+--+,+++-)$ \\
$(++-+,++-+)$ & $(++-+,++-+)$ & $(+--+,++-+)$ & $(+--+,++-+)$ \\
$(++-+,++--) $ & $(++-+,++--) $ & $(+--+,++--) $ & $(+--+,++--) $ \\
$(++-+,+-++)$ & $(++-+,+-++)$ & $(+--+,+-++)$ & $(+--+,+-++)$ \\
$(++-+,+-+-)$ & $(++-+,+-+-)$ & $(+--+,+-+-)$ & $(+--+,+-+-)$ \\
$(++-+,+--+)$ & $(++-+,+--+)$ & $(+--+,+--+)$ & $(+--+,+--+)$ \\
$(++-+,+---)$ & $(++-+,+---)$ & $(+--+,+---)$ & $(+--+,+---)$ \\
$(++-+,-+++)$ & $(+--+,-+++)$ & $(+--+,-+++)$ & $(++-+,-+++)$ \\
$(++-+,-++-)$ & $(+--+,-++-)$ & $(+--+,-++-)$ & $(++-+,-++-)$ \\
$(++-+,-+-+)$ & $(+--+,-+-+)$ & $(+--+,-+-+)$ & $(++-+,-+-+)$ \\
$(++-+,-+--)$ & $(+--+,-+--)$ & $(+--+,-+--)$ & $(++-+,-+--)$ \\
$(++-+,--++)$ & $(+--+,--++)$ & $(+--+,--++)$ & $(++-+,--++)$ \\
$(++-+,--+-)$ & $(+--+,--+-)$ & $(+--+,--+-)$ & $(++-+,--+-)$ \\
$(++-+,---+)$ & $(+--+,---+)$ & $(+--+,---+)$ & $(++-+,---+)$ \\
$(++-+,----)$ & $(+--+,----)$ & $(+--+,----)$ & $(++-+,----)$ \\
$(++--,++++)$ & $(++--,++++)$ & $(+---,++++)$ & $(+---,++++)$ \\
$(++--,+++-)$ & $(++--,+++-)$ & $(+---,+++-)$ & $(+---,+++-)$ \\
$(++--,++-+)$ & $(++--,++-+)$ & $(+---,++-+)$ & $(+---,++-+)$ \\
$(++--,++--) $ & $(++--,++--) $ & $(+---,++--) $ & $(+---,++--) $ \\
$(++--,+-++)$ & $(++--,+-++)$ & $(+---,+-++)$ & $(+---,+-++)$ \\
$(++--,+-+-)$ & $(++--,+-+-)$ & $(+---,+-+-)$ & $(+---,+-+-)$ \\
$(++--,+--+)$ & $(++--,+--+)$ & $(+---,+--+)$ & $(+---,+--+)$ \\
$(++--,+---)$ & $(++--,+---)$ & $(+---,+---)$ & $(+---,+---)$ \\
$(++--,-+++)$ & $(+---,-+++)$ & $(+---,-+++)$ & $(++--,-+++)$ \\
$(++--,-++-)$ & $(+---,-++-)$ & $(+---,-++-)$ & $(++--,-++-)$ \\
$(++--,-+-+)$ & $(+---,-+-+)$ & $(+---,-+-+)$ & $(++--,-+-+)$ \\
$(++--,-+--)$ & $(+---,-+--)$ & $(+---,-+--)$ & $(++--,-+--)$ \\
$(++--,--++)$ & $(+---,--++)$ & $(+---,--++)$ & $(++--,--++)$ \\
$(++--,--+-)$ & $(+---,--+-)$ & $(+---,--+-)$ & $(++--,--+-)$ \\
$(++--,---+)$ & $(+---,---+)$ & $(+---,---+)$ & $(++--,---+)$ \\
$(++--,----)$ & $(+---,----)$ & $(+---,----)$ & $(++--,----)$ \\
\end{tabular}
\end{ruledtabular}
\end{table*}

\begin{table*}[p]
\caption{\label{tab:b_group_operations_2} Full mapping of a \(B\)-group operation on one leaf of the \(4\)-qubit linear state (Part II).}
\renewcommand{\arraystretch}{0.90}
\begin{ruledtabular}
\begin{tabular}{llll}
\multicolumn{2}{c}{Part II-a} & \multicolumn{2}{c}{Part II-b} \\
\hline
Initial state & After B group operation &
Initial state & After B group operation \\
\hline
$(-+++,++++)$ & $(-+++,+-++)$ & $(--++,++++)$ & $(--++,+-++)$ \\
$(-+++,+++-)$ & $(-+++,+-+-)$ & $(--++,+++-)$ & $(--++,+-+-)$ \\
$(-+++,++-+)$ & $(-+++,+--+)$ & $(--++,++-+)$ & $(--++,+--+)$ \\
$(-+++,++--) $ & $(-+++,+---) $ & $(--++,++--) $ & $(--++,+---) $ \\
$(-+++,+-++)$ & $(-+++,++++)$ & $(--++,+-++)$ & $(--++,++++)$ \\
$(-+++,+-+-)$ & $(-+++,+++-)$ & $(--++,+-+-)$ & $(--++,+++-)$ \\
$(-+++,+--+)$ & $(-+++,++-+)$ & $(--++,+--+)$ & $(--++,++-+)$ \\
$(-+++,+---)$ & $(-+++,++--) $ & $(--++,+---)$ & $(--++,++--) $ \\
$(-+++,-+++)$ & $(--++,--++)$ & $(--++,-+++)$ & $(-+++,--++)$ \\
$(-+++,-++-)$ & $(--++,--+-)$ & $(--++,-++-)$ & $(-+++,--+-)$ \\
$(-+++,-+-+)$ & $(--++,---+)$ & $(--++,-+-+)$ & $(-+++,---+)$ \\
$(-+++,-+--)$ & $(--++,----)$ & $(--++,-+--)$ & $(-+++,----)$ \\
$(-+++,--++)$ & $(--++,-+++)$ & $(--++,--++)$ & $(-+++,-+++)$ \\
$(-+++,--+-)$ & $(--++,-++-)$ & $(--++,--+-)$ & $(-+++,-++-)$ \\
$(-+++,---+)$ & $(--++,-+-+)$ & $(--++,---+)$ & $(-+++,-+-+)$ \\
$(-+++,----)$ & $(--++,-+--)$ & $(--++,----)$ & $(-+++,-+--)$ \\
$(-++-,++++)$ & $(-++-,+-++)$ & $(--+-,++++)$ & $(--+-,+-++)$ \\
$(-++-,+++-)$ & $(-++-,+-+-)$ & $(--+-,+++-)$ & $(--+-,+-+-)$ \\
$(-++-,++-+)$ & $(-++-,+--+)$ & $(--+-,++-+)$ & $(--+-,+--+)$ \\
$(-++-,++--) $ & $(-++-,+---) $ & $(--+-,++--) $ & $(--+-,+---) $ \\
$(-++-,+-++)$ & $(-++-,++++)$ & $(--+-,+-++)$ & $(--+-,++++)$ \\
$(-++-,+-+-)$ & $(-++-,+++-)$ & $(--+-,+-+-)$ & $(--+-,+++-)$ \\
$(-++-,+--+)$ & $(-++-,++-+)$ & $(--+-,+--+)$ & $(--+-,++-+)$ \\
$(-++-,+---)$ & $(-++-,++--) $ & $(--+-,+---)$ & $(--+-,++--) $ \\
$(-++-,-+++)$ & $(--+-,--++)$ & $(--+-,-+++)$ & $(-++-,--++)$ \\
$(-++-,-++-)$ & $(--+-,--+-)$ & $(--+-,-++-)$ & $(-++-,--+-)$ \\
$(-++-,-+-+)$ & $(--+-,---+)$ & $(--+-,-+-+)$ & $(-++-,---+)$ \\
$(-++-,-+--)$ & $(--+-,----)$ & $(--+-,-+--)$ & $(-++-,----)$ \\
$(-++-,--++)$ & $(--+-,-+++)$ & $(--+-,--++)$ & $(-++-,-+++)$ \\
$(-++-,--+-)$ & $(--+-,-++-)$ & $(--+-,--+-)$ & $(-++-,-++-)$ \\
$(-++-,---+)$ & $(--+-,-+-+)$ & $(--+-,---+)$ & $(-++-,-+-+)$ \\
$(-++-,----)$ & $(--+-,-+--)$ & $(--+-,----)$ & $(-++-,-+--)$ \\
$(-+-+,++++)$ & $(-+-+,+-++)$ & $(---+,++++)$ & $(---+,+-++)$ \\
$(-+-+,+++-)$ & $(-+-+,+-+-)$ & $(---+,+++-)$ & $(---+,+-+-)$ \\
$(-+-+,++-+)$ & $(-+-+,+--+)$ & $(---+,++-+)$ & $(---+,+--+)$ \\
$(-+-+,++--) $ & $(-+-+,+---) $ & $(---+,++--) $ & $(---+,+---) $ \\
$(-+-+,+-++)$ & $(-+-+,++++)$ & $(---+,+-++)$ & $(---+,++++)$ \\
$(-+-+,+-+-)$ & $(-+-+,+++-)$ & $(---+,+-+-)$ & $(---+,+++-)$ \\
$(-+-+,+--+)$ & $(-+-+,++-+)$ & $(---+,+--+)$ & $(---+,++-+)$ \\
$(-+-+,+---)$ & $(-+-+,++--) $ & $(---+,+---)$ & $(---+,++--) $ \\
$(-+-+,-+++)$ & $(---+,--++)$ & $(---+,-+++)$ & $(-+-+,--++)$ \\
$(-+-+,-++-)$ & $(---+,--+-)$ & $(---+,-++-)$ & $(-+-+,--+-)$ \\
$(-+-+,-+-+)$ & $(---+,---+)$ & $(---+,-+-+)$ & $(-+-+,---+)$ \\
$(-+-+,-+--)$ & $(---+,----)$ & $(---+,-+--)$ & $(-+-+,----)$ \\
$(-+-+,--++)$ & $(---+,-+++)$ & $(---+,--++)$ & $(-+-+,-+++)$ \\
$(-+-+,--+-)$ & $(---+,-++-)$ & $(---+,--+-)$ & $(-+-+,-++-)$ \\
$(-+-+,---+)$ & $(---+,-+-+)$ & $(---+,---+)$ & $(-+-+,-+-+)$ \\
$(-+-+,----)$ & $(---+,-+--)$ & $(---+,----)$ & $(-+-+,-+--)$ \\
$(-+--,++++)$ & $(-+--,+-++)$ & $(----,++++)$ & $(----,+-++)$ \\
$(-+--,+++-)$ & $(-+--,+-+-)$ & $(----,+++-)$ & $(----,+-+-)$ \\
$(-+--,++-+)$ & $(-+--,+--+)$ & $(----,++-+)$ & $(----,+--+)$ \\
$(-+--,++--) $ & $(-+--,+---) $ & $(----,++--) $ & $(----,+---) $ \\
$(-+--,+-++)$ & $(-+--,++++)$ & $(----,+-++)$ & $(----,++++)$ \\
$(-+--,+-+-)$ & $(-+--,+++-)$ & $(----,+-+-)$ & $(----,+++-)$ \\
$(-+--,+--+)$ & $(-+--,++-+)$ & $(----,+--+)$ & $(----,++-+)$ \\
$(-+--,+---)$ & $(-+--,++--) $ & $(----,+---)$ & $(----,++--) $ \\
$(-+--,-+++)$ & $(----,--++)$ & $(----,-+++)$ & $(-+--,--++)$ \\
$(-+--,-++-)$ & $(----,--+-)$ & $(----,-++-)$ & $(-+--,--+-)$ \\
$(-+--,-+-+)$ & $(----,---+)$ & $(----,-+-+)$ & $(-+--,---+)$ \\
$(-+--,-+--)$ & $(----,----)$ & $(----,-+--)$ & $(-+--,----)$ \\
$(-+--,--++)$ & $(----,-+++)$ & $(----,--++)$ & $(-+--,-+++)$ \\
$(-+--,--+-)$ & $(----,-++-)$ & $(----,--+-)$ & $(-+--,-++-)$ \\
$(-+--,---+)$ & $(----,-+-+)$ & $(----,---+)$ & $(-+--,-+-+)$ \\
$(-+--,----)$ & $(----,-+--)$ & $(----,----)$ & $(-+--,-+--)$ \\
\end{tabular}
\end{ruledtabular}
\end{table*}

\end{document}